\documentclass[modern]{aastex63}
\turnoffediting
\pdfoutput=1
\usepackage{amsmath}
\usepackage{hyperref}

\received{XXXX XX, 2021}
\revised{XXXX XX, 2021}
\accepted{XXXX XX, 2021}
\submitjournal{XX}
\shorttitle{de-trending with \texttt{unpopular}}
\shortauthors{Hattori et al.}

\sloppypar
\begin{document}

\title{The \texttt{unpopular} Package: A Data-driven Approach to De-trend \edit1{TESS} \newline Full Frame Image Light Curves}

\newcommand{\affnyuad}{New York University Abu Dhabi,
PO Box 129188, Abu Dhabi, United Arab Emirates}
\newcommand{\affcca}{Center for Computational Astrophysics, Flatiron Institute, 162 Fifth Ave, New York, NY 10010, USA}
\newcommand{\affccpp}{Center for Cosmology and Particle Physics, Department of Physics, New York University, 726 Broadway, New York, NY 10003, USA}
\newcommand{\affmpia}{Max-Planck-Institut f\"ur Astronomie, K\"onigstuhl 17, D-69117 Heidelberg, Germany}
\newcommand{\affmpiis}{Max-Planck-Institut f\"ur Intelligente Systeme, Max-Planck-Ring 4, D-72076 T\"ubingen, Germany}
\newcommand{\unsw}{School of Physics, University of New South Wales, Sydney, NSW 2052, Australia}
\newcommand{\affamnh}{Department of Astrophysics, American Museum of Natural History, 200 Central Park West, New York, NY 10024, USA}
\newcommand{\columbia}{Department of Astronomy, Columbia University, 550 West 120th Street, New York, NY 10027, USA}

\correspondingauthor{Soichiro~Hattori}
\email{soichiro.hattori@columbia.edu}

\author[0000-0002-0842-863X]{Soichiro~Hattori}
\affiliation{\columbia}

\author[0000-0002-9328-5652]{Daniel~Foreman-Mackey}
\affiliation{\affcca}

\author[0000-0003-2866-9403]{David~W.~Hogg}
\affiliation{\affcca}
\affiliation{\affmpia}
\affiliation{\affccpp}

\author[0000-0001-7516-8308]{Benjamin~T.~Montet}
\affiliation{\unsw}

\author[0000-0003-4540-5661]{Ruth Angus}
\affiliation{\affamnh}
\affiliation{\affcca}

\author[0000-0001-9227-8349]{T. A. Pritchard}
\affiliation{\affccpp}

\author[0000-0002-2792-134X]{Jason L.~Curtis}
\affiliation{\columbia}
\affiliation{\affamnh}

\author[0000-0002-8177-0925]{Bernhard Sch\"olkopf}
\affiliation{\affmpiis}

\begin{abstract}\noindent
The majority of observed pixels on the Transiting Exoplanet Survey Satellite (TESS) are delivered in the form of full frame images (FFI). However, the FFIs contain systematic effects such as pointing jitter and scattered light from the Earth and Moon that must be removed \edit1{(i.e., ``de-trended")} before downstream analysis. We present \texttt{unpopular}, an open-source Python package to \edit1{obtain} de-trended TESS FFI light curves \edit1{optimized for variable sources}. \edit1{\texttt{unpopular} implements a variant of the causal pixel model to remove systematics and allows for simultaneous fitting with a polynomial component to capture non-transit astrophysical variations, such as supernova signals or stellar variability, that tend to be removed in techniques optimized for exoplanet detection.} We validate our method by de-trending different sources (e.g., supernova, tidal disruption event (TDE), exoplanet-hosting star, fast-rotating star) and comparing our light curves to those obtained by other pipelines when appropriate. \edit1{Our supernova and TDE light curves are visually similar to those obtained by others using the ISIS image subtraction package, indicating that \texttt{unpopular} can be used to extract multi-sector light curves by preserving astrophysical signals on timescales of a TESS sector ($\sim{27}$ days). We note that our method contains tuning parameters that are currently set heuristically, and that the optimal set of tuning parameters will likely depend on the particular signal the user is interested in obtaining.} 
The \texttt{unpopular} source code and tutorials are freely available online.
\end{abstract}


\section{Introduction} \label{sec:intro}

\begin{figure}
    \centering
    \includegraphics[width=\textwidth]{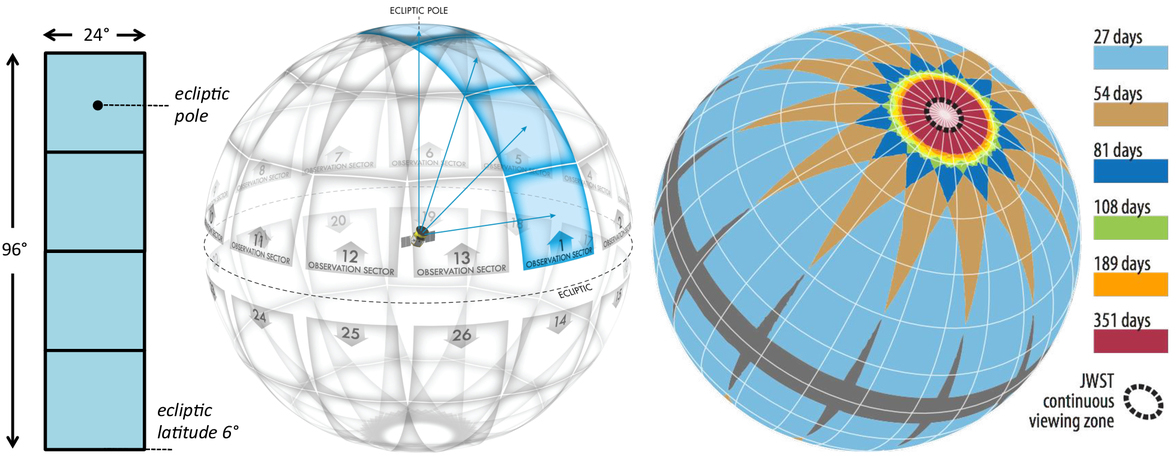}
    \caption{The observation strategy of TESS (Figure 7 from \citet{ricker2014} licensed under \href{https://creativecommons.org/licenses/by/4.0/}{CC BY 4.0}). The left panel shows the field of view of a single TESS sector created by its four cameras. The middle panel shows how TESS divides and observes the sky in 26 partially overlapping sectors. The right panel shows how the overlap between sectors leads to certain ares of they sky being observed for longer durations.}
    \label{fig:tess_obs}
\end{figure}

During its 2-year primary mission, the \textsl{Transiting Exoplanet Survey Satellite (TESS)} \citep{ricker2014} observed approximately 200,000 preselected main-sequence stars of spectral types F5 to M5 at 2-minute cadence. The majority of these stars were chosen to satisfy one of the primary science requirements of detecting small $(R_p < 4R_\oplus)$ exoplanets around nearby bright \edit1{($I_c \lesssim 10\text{-}13$ where $I_c$ is the magnitude in the Johnson-Cousins 700-900 nm bandpass filter)} stars, sufficiently bright for further characterization with ground-based observations \citep{ricker2014}. 
In addition to the preselected stars, during each of its 26 observation sectors, where an observation sector is a $24\degr \times 96\degr$ \edit1{strip} of the sky \edit1{continuously} monitored for approximately 27 days \edit1{with TESS's four cameras} (see Figure \ref{fig:tess_obs}), TESS observed ${\sim} 10^6$ bright sources $(I_c < 14\text{-}15)$ in \edit1{a series of} full frame images (FFIs) taken at 30-minute cadence \citep{ricker2014}. \edit1{As shown in Figure \ref{fig:tess_obs}, some sources were monitored for longer than 27 days as the sectors overlap at high ecliptic latitudes; in particular, the $24\degr$-diameter circular region around the ecliptic poles were monitored for ${\sim}1$ year and are called the Continuous-Viewing Zones (CVZs).}

\edit1{With approximately 20 million targets in the FFI data \citep{ricker2014} for which relative photometry with 1\% photometric precision can be achieved, the FFI data not only expand the search for exoplanet transits beyond the preselected stars, such as the detection of TOI-172b \citep{rodriguez2019}, TOI-942b, and TOI-942c \citep{carleo2021, zhou2021}, but also provide a rich dataset to pursue scientific investigations on a variety of astronomical phenomena (e.g., supernovae, asteroids, flaring stars, eclipsing binaries, tidal disruption events, variable stars, microlensing events). Some examples include studying the early-time behavior of transient events, such as the Type-Ia supernova ASASSN-18tb/SN 2018fhw \citep{vallely2019} and the tidal disruption event ASASSN-19bt/AT 2019ahk \citep{holoien2019}. The FFI data has also shown promising potential for detecting distant Solar System objects \citep{holman2019, payne2019}, where \citet{rice2020} were recently able to successfully recover the signals of three trans-Neptunian objects and \citet{bernardinelli2021} identified cometary activity in TESS images of the ``mega-comet'' C/2014 UN271 (Bernardinelli-Bernstein).}

\begin{figure}
    \centering
    \includegraphics[width=\textwidth]{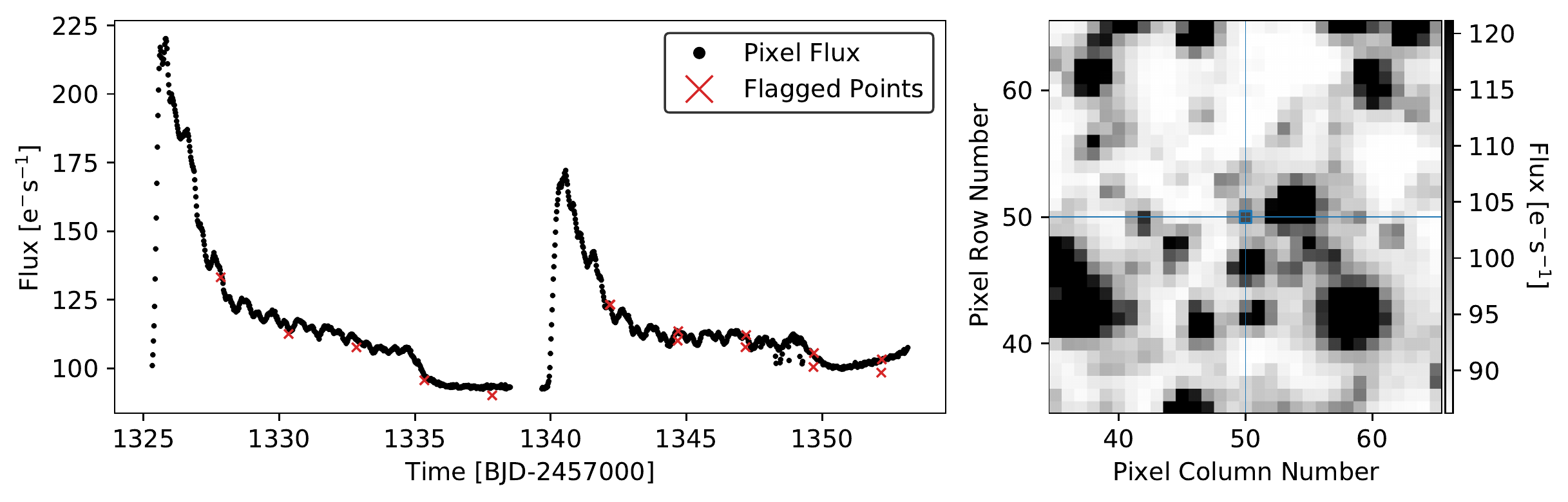}
    \caption{The systematic effects in the FFI data. In the left panel we show the Sector 1 raw light curve of the central pixel of the Type-Ia supernova ASASSN-18tb/SN 2018fhw. The sudden flux increases followed by a gradual decline at the beginning and middle of the light curve are caused by the scattered light from the Earth and Moon. The gap in the middle is when the data downlink occurred. The flagged points (red) indicate when the momentum dumps, a period of reduced pointing stability due to the spacecraft firing its thrusters, occurred. The high pointing jitter between 1348-1350 
    days was caused by an improper spacecraft configuration. In the right panel we show a $40 \times 40$ pixel FFI cutout centered on ASASSN-18tb from Sector 1. The median value for each pixel is shown, and the color bar is calibrated to the 10-90\% range of flux values. The central pixel of ASASSN-18tb is located where the two blue lines intersect and is surrounded by a blue square.
    }
    
    \label{fig:scattered_light_asassn18tb}
\end{figure}

\edit1{While the calibrated FFIs released by the TESS Science Processing Operations Center (SPOC) pipeline have been corrected for systematics such as CCD effects (e.g., bias, dark current) and cosmic rays \citep{jenkins2016}, further processing is required to remove additional observational systematics\footnote{\url{https://archive.stsci.edu/missions/tess/doc/TESS_Instrument_Handbook_v0.1.pdf}} (e.g., scattered light) when extracting light curves.} Similar to the Kepler spacecraft \citep{stumpe2012}, some causes for these systematics include periods of reduced spacecraft pointing accuracy and changes in detector sensitivity due to temperature variations. \edit1{The} dominant systematic effect for TESS is scattered light from the Earth and Moon entering the cameras and increasing the measured flux of a given pixel (Figure \ref{fig:scattered_light_asassn18tb}). This effect is caused by the spacecraft's unique eccentric 13.7-day orbit, half the orbital period of the Moon, and usually occurs preceding or following a data downlink (which happens after each orbit) when the spacecraft is close to Earth\edit1{\footnote{A more complete explanation is provided on \url{https://tess.mit.edu/observations/scattered-light/} and the scattered light is analyzed in the data release notes (\url{https://archive.stsci.edu/tess/tess_drn.html})}}.

Fortunately, many methods to remove systematics from light curves (i.e., ``de-trending") were developed during the Kepler/K2 era and serve as foundations for developing TESS de-trending methods. Correcting for systematics became particularly important during the K2 mission as the spacecraft was operating with reduced pointing accuracy compared to the original Kepler mission.

\edit1{Some} of these methods relied on the idea that flux variations that correlated with the source position on the CCD were likely to be systematic effects, an idea previously used to remove systematic effects in the photometric time series data from the Spitzer Space Telescope \citep{charbonneau2005, knutson2008, ballard2010, stevenson2012}. Using this ``self-flat-fielding (SFF)" approach, \citet{vanderburg2014}, \citet{armstrong2015}, \edit1{and \citet{lund2015}} produced K2 light curves corrected for spacecraft motion, the dominant systematic effect. \citet{aigrain2015, aigrain2016} and \citet{crossfield2015} modeled the spacecraft's \edit1{net} systematics as a Gaussian process, with pointing measurements as the input, and obtained de-trended light curves by subtracting the prediction of that model from the observed flux.

\edit1{Other methods, which trace their conceptual heritage to the Trend Filtering Algorithm (TFA) developed by \citet{kovacs2005}, relied on the observation that systematic effects would likely be shared by many distant pixels across the detectors, and modeled the systematics as a linear combination of some basis vectors (i.e., regressors) generated by the photometric data of other pixels.
Some examples include the official Kepler pipeline's Presearch Data Conditioning module (PDC; \citealt{smith2012, stumpe2012, stumpe2014}), the ``eigen light curve" approach taken by \citet{foreman-mackey2015}, the \texttt{EVEREST} pipeline \citep{luger2016, luger2018}, and the \textsl{Causal Pixel Model (CPM)}-based pipelines \citep{wang2016, wang2017, poleski2019}. 
As further discussed in \S\ref{sec:cpm}, some differences between them include their choice of regressors and whether they de-trend at the aperture photometry level or at the individual pixel level. There are also de-trending pipelines that incorporate both approaches; the pipeline developed by \citet{huang2015} performed an initial de-trending step similar to SFF and then subsequently used the TFA. As both TESS and Kepler share a similar observing strategy of wide-field continuous observations, these data-driven de-trending approaches can be applied to TESS FFI data with minimal conceptual modifications.}

\edit1{We therefore build upon the \textsl{CPM} \citep{wang2016} and \textsl{CPM Difference Imaging} \citep{wang2017} pipelines and present \texttt{unpopular}, an open-source Python package for extracting de-trended TESS FFI light curves by removing common (i.e., popular) trends at the individual pixel-level data. \texttt{unpopular} was designed to obtain light curves for variable sources by implementing a variant of \textsl{CPM} to remove systematics and allowing for simultaneous fitting with a polynomial component to capture the non-transit astrophysical variations, such as supernova signals and stellar variability, that tend to be removed in techniques optimized for exoplanet detection.} 

The \texttt{unpopular} package will complement existing open-source packages for extracting TESS FFI light curves. Two such popular packages are \texttt{eleanor} \citep{feinstein2019} and \texttt{lightkurve} \citep{barentsen2020}. \edit1{\texttt{eleanor} is a package designed to generate light curves optimized for exoplanet searches and \texttt{lightkurve} is a general-use package that allows for user-friendly interfacing and de-trending of the FFI data. In comparison, \texttt{unpopular} was designed to be a useful tool for efficiently obtaining light curves of variable sources by preserving long-term astrophysical variations. We show that it can successfully recover supernova and tidal disruption event (TDE) light curves. We note that a beta version of the \texttt{unpopular} package was used to obtain the rotation periods of stars in the 700-Myr-old Coma Berenices (Coma Ber) open cluster \citep{singh2021}, the 38-Myr-old $\delta$ Lyra cluster \citep{Bouma2021}, and in the $\sim$150-Myr-old Theia~456 stellar stream \citep{Andrews2021}.}

This paper is organized as follows. In \S\ref{sec:cpm}, we discuss the \textsl{CPM} method and the mathematical formulation behind the \texttt{unpopular} de-trending method. In \S\ref{sec:unpopular}, we describe the steps involved in creating light curves using the \texttt{unpopular} package and show an example of extracting a supernova light curve. In \S\ref{sec:results_examples}, we present \edit1{additional} de-trended FFI light curves. In \S\ref{sec:conclusion}, we discuss certain aspects of the method and summarize our work.  

\section{Causal Pixel Model} \label{sec:cpm}

In this section we describe the concepts behind the \textsl{CPM} (and its variant \textsl{CPM Difference Imaging}) method and discuss the modifications we have made in our \texttt{unpopular} implementation. As mentioned in \S\ref{sec:intro}, \textsl{CPM} is within the family of methods that attempts to model the systematics as a combination of some chosen set of \edit1{regressors} \citep[e.g.,][]{smith2012, stumpe2012, foreman-mackey2015, luger2016, luger2018, wang2016, wang2017, poleski2019}. \edit1{These methods all de-trend based on the realization that for a given observation in wide-field photometric surveys, such as Kepler and TESS, systematic trends are shared by many \textsl{different} sources on the focal plane (i.e., light curve variations that are common across distance pixels are likely to be systematics) \citep{kovacs2005}. As such, the regressors are generated from pixels \textsl{outside the target source's aperture}. \texttt{EVEREST} \citep{luger2016, luger2018} is unique among these methods in that in addition to this de-trending step, it also performs ``pixel-level decorrelation" (PLD), a de-trending technique initially developed by \citet{deming2015} for Spitzer data that uses pixels \textsl{within the aperture} as its regressors. In PLD, the regressors are generated by normalizing each of the pixels by the aperture-summed flux, effectively removing the astrophysical signal and leaving a set of basis vectors that only contain signals that are different across the aperture. As \texttt{unpopular} employs \textsl{CPM}, a method that relies on shared systematics \textsl{across multiple sources}, the regressors are only chosen from outside the aperture.}

\begin{figure}
    \centering
    \includegraphics[width=0.6\textwidth]{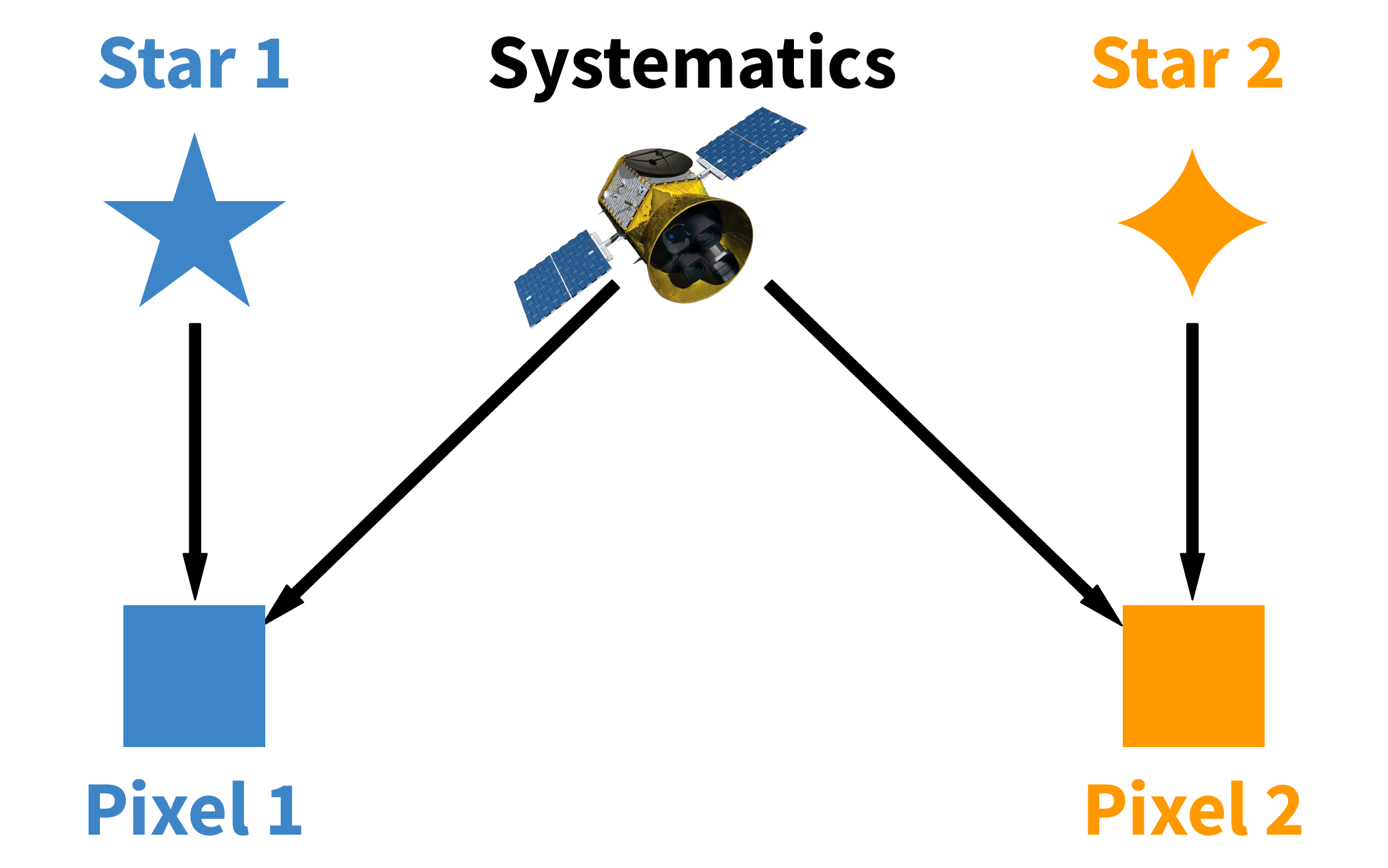}
    \caption{Diagram motivating the half-sibling regression method. The quantities in the top row (\textbf{Star 1, Systematics, Star 2}) are unobserved while the quantities in the bottom row (\textbf{Pixel 1, Pixel 2}) are observed. The two pixels are on the same CCD and \textsl{separated by at least several times the size of the point spread function to ensure that no flux from one of the stars is simultaneously illuminating both pixels.}
    \textbf{Pixel 1} and \textbf{Pixel 2} can be thought of as ``half-siblings" as they both share a common parent (\textbf{Systematics}).
    \textsl{TESS Spacecraft Image Credit: NASA's Goddard Space Flight Center}}
    \label{fig:hsr}
\end{figure}

\edit1{\citet{scholkopf2016} provide a formal mathematical framework for de-trending based on shared systematics and refer to it as ``half-sibling regression". We briefly describe the concept in the context of TESS and refer the reader to \citet{scholkopf2016} for a detailed mathematical analysis. The half-sibling relationship refers to how, as shown in Figure \ref{fig:hsr}, two pixels on the same CCD illuminated by \textsl{different} stars, separated by more than a few times the size of the PSF to ensure that neither star is simultaneously illuminating both pixels, share \textsl{one common parent}: the TESS systematics. Given that two stars physically separated by light-years are unlikely to vary in a synchronized way, we can infer that contemporaneous pixel light curve variations are caused by the TESS systematics. \textsl{In other words, while one of the light curves can be used to predict the systematic variations in the other light curve, as those come from the same parent (i.e., TESS systematics) and are shared across both pixels, it cannot predict the astrophysical variations, as those come from different parents (i.e., two different stars) and are unique to each light curve.} We can therefore model the systematics in a given light curve using other light curves from distant pixels.}

\edit1{In practice multiple light curves are used as regressors to increase the flexibility of the model; the original \textsl{CPM} method used 4000 light curve regressors. While not all systematics can be completely removed from the target light curve, this data-driven approach can efficiently remove significant common systematic effects. The half-sibling regression framework has also been recently applied to improve direct detection of exoplanets in the area of High-Contrast Imaging \citep{samland2021, gebhard2020}.}

While \edit1{these methods share the same principle, as mentioned in \S\ref{sec:intro}} one key difference between them is the choice of regressors.
In the official Kepler pipeline's Presearch Data Conditioning module (PDC; \citealt{smith2012, stumpe2012, stumpe2014}), the module responsible for correcting systematics, the regressors were the top eight ``Cotrending Basis Vectors" (CBVs). 
The CBVs were different for each of the \edit1{84 Kepler output channels, where a channel read out half of the set of pixels for each of the 42 CCDs on the photometer,} and they were obtained by performing Singular Value Decomposition (SVD) on a set of highly correlated and quiet light curves that were then ranked by their singular values. \edit1{From} version 8.2 of the Kepler pipeline \edit1{``multiscale error correction" was implemented, where} the above process is performed after first \edit1{using an overcomplete discrete wavelet transform to decompose a single light curve into multiple time-series, each time-series only containing variations within a range of frequencies (i.e., band-splitting) \citep{stumpe2014}.}. For each of these time-series of a particular scale, an accompanying set of CBVs is generated and used as regressors in the de-trending step. \citet{foreman-mackey2015} took a similar approach by performing Principal Component Analysis (PCA) on a set of K2 Campaign 1 light curves and used the top 150 principal components, referred to as ``eigen light curves", as the regressors. \texttt{EVEREST} \citep{luger2016, luger2018} uses products of the normalized pixel fluxes in a chosen K2 target's aperture as the regressors \edit1{in the PLD step and normalized pixel fluxes from neighboring stars when de-trending based on shared systematics}. \edit1{The \textsl{CPM}-based methods \citep{wang2016, wang2017, poleski2019}} and the \texttt{unpopular} implementation use the light curves of many other pixels illuminated by distant sources on the same CCD as the regressors.

One subtle but significant aspect that sets \textsl{CPM}-based methods apart from the rest of these methods is that it works exclusively at the pixel level. Kepler PDC and \citet{foreman-mackey2015} work exclusively at the aperture photometry level, where both the quantity being modeled (i.e., the regressand) and the regressors are simple aperture photometry (SAP)\footnote{In simple aperture photometry the values of the pixels within a particular aperture are summed together.} light curves (technically the regressors are obtained by performing dimensionality reduction on SAP light curves). Aperture photometry is advantageous as individual pixel-level variations caused by systematics, such as pointing drift, can be significantly reduced by choosing an appropriately large aperture. However, particularly for TESS, where choosing a large aperture risks contamination from nearby sources given the relatively large pixels ($21$ arcseconds), working at the pixel level can yield better results. \texttt{EVEREST} incorporates pixel-level variations by generating the regressors from individual pixels' flux measurements, although the regressand is still an SAP light curve. \textsl{CPM}-based methods go one step further where both the regressand and the regressors are individual pixel's flux measurements. While a downside to this decision is that each pixel within an aperture must be \edit1{processed} separately, the computational efficiency of our implementation makes this problem tractable. 

A significant concern for these linear models is the possibility of \textsl{overfitting} given their flexibility. That is, since we are fitting a linear combination of multiple regressors to a single time series (i.e., a pixel or aperture light curve), there is a possibility that the systematics model will not only capture the systematics but also variations from the astrophysical signal. One approach to mitigate overfitting is to reduce the flexibility of the model by using fewer regressors; this approach was adopted in the Kepler PDC module \citep{smith2012, stumpe2012}. \citet{foreman-mackey2015} prevents the systematics model from overfitting transits, the signal of interest in that study, by simultaneously fitting a transit model to the raw light curves. Another approach is regularization, in which an additional term is added to the objective function to control the behavior of the model and prevent overfitting; this approach was adopted in the \textsl{CPM} methods \citep{wang2016, wang2017, poleski2019}. \texttt{EVEREST 2.0} \citep{luger2018} combines both regularization and the simultaneous fitting of a Gaussian process to capture the astrophysical variability. The current approach in \texttt{unpopular} is somewhat similar to \texttt{EVEREST 2.0} as we also use regularization and allow for simultaneous fitting with an additional model to capture astrophysical variability. However, instead of modeling the astrophysical variability with a Gaussian process as done in \texttt{EVEREST 2.0}, we use a simple polynomial model for computational efficiency. 

Modifications to the original \textsl{CPM} method have been made in the \texttt{unpopular} implementation to reflect a change in the scientific objective. In the original \textsl{CPM} implementation, the objective was to detect exoplanet transits in the de-trended light curves. The model was tuned to specifically preserve transit signals, and the intrinsic stellar variability was removed by adding an autoregressive component. As the objective with \texttt{unpopular} is to obtain de-trended light curves for a variety of astrophysical sources, we prioritize preserving non-transit signals including stellar variability. Therefore, we \edit1{omit the} autoregressive component and instead include the previously mentioned optional polynomial component. \edit1{This change is to improve our sensitivity to transient sources with long-term variability and is not a statement that we believe a polynomial approach is superior to other approaches (e.g., autoregressive approaches).}


\subsection{Ordinary Least Squares}\label{sec:ols}
We will first write the mathematical formulation for the \textsl{ordinary least squares} (OLS) method. In the following subsection \S\ref{sec:ridge_regression} we discuss \textsl{ridge regression}, an extension to OLS that incorporates regularization and is the method used in \texttt{unpopular}.
We use the notation where lowercase upright fonts in bold ($\mathbf{y}$) are row or column vectors and uppercase upright fonts in bold ($\mathbf{X}$) are matrices. 

\edit1{In \texttt{unpopular} we model a single pixel's light curve as a linear combination of $L$ predictor pixel light curves and an optional $P$-degree polynomial model, to capture the systematics and long-term astrophysical trends, respectively. An offset term is also added to capture the mean value of the light curve.}
\edit1{Our linear model can be written as 
\begin{equation} \label{eqn:basic_linear_equation}
    \mathbf{y} = \mathbf{X} \cdot \mathbf{w} + \mathbf{e}
\end{equation}
where $\mathbf{y, e}$ are $N \times 1$ column vectors containing the $N$ flux measurements, $y_i$, and the associated measurement uncertainties, $e_i$, respectively. When including the optional polynomial model the $N \times (L+P+1)$ \textsl{design matrix} $\mathbf{X}$ is 
\begin{equation}
    \mathbf{X} = 
    \begin{bmatrix}
    \mathbf{X_\mathrm{sys}} & \mathbf{X_\mathrm{poly}} & \mathbf{1} 
    \end{bmatrix} = 
    \begin{bmatrix}
    l_{1,1} & l_{1,2} & \cdots & l_{1,L} & t^{P}_1 & t^{P-1}_1 & \cdots & t_1 & 1\\ 
    l_{2,1} & l_{2,2} & \cdots & l_{2,L} & t^{P}_2 & t^{P-1}_2 & \cdots & t_2 & 1\\
    \vdots & \vdots & \vdots & \vdots & \vdots & \vdots & \vdots & \vdots & \vdots\\ 
    l_{N,1} & l_{N,2} & \cdots & l_{N,L} & t^{P}_N & t^{P-1}_N & \cdots & t_N & 1\\
    \end{bmatrix},
\end{equation}
where $\mathbf{X_\mathrm{sys}}$ is an $N \times L$ matrix of the $L$ predictor pixel light curve regressors, $\mathbf{X_\mathrm{poly}}$ is an $N \times P$ matrix of the the $P$ polynomial regressors, and $1$ is simply an $N \times 1$ column vector of ones. The $(L+P+1) \times 1$ column vector $\mathbf{w}$ contains the coefficients for each regressor. As shown in \ref{appendix:proof_ols}, the OLS solution, obtained by minimizing the sum-of-squared-residuals objective function
\begin{equation}
    S(\mathbf{w}) = \lVert \mathbf{y} - \mathbf{X} \cdot \mathbf{w} \rVert_2^2= (\mathbf{y} - \mathbf{X} \cdot \mathbf{w})^\top \cdot (\mathbf{y} - \mathbf{X} \cdot \mathbf{w}),
\end{equation}
is given by
\begin{equation}
    \mathbf{\hat{w}} = (\mathbf{X^\top \cdot X})^{-1} \cdot (\mathbf{X^\top \cdot y})
\end{equation}
and is equivalent to maximizing the likelihood for the model coefficients when treating the uncertainties $e_i$ as being Gaussian-distributed \citep{hogg2010}.} 

\edit1{Once the vector of coefficients $\mathbf{\hat{w}}$ has been calculated, we can construct the best estimate for the model $\mathbf{\hat{m}}$ as
\begin{equation}
    \mathbf{\hat{m}} = \mathbf{X} \cdot \mathbf{\hat{w}}.
\end{equation}
When the optional polynomial component is used, the model $\mathbf{\hat{m}}$ can be separated into three parts
\begin{equation}
    \mathbf{\hat{m}} = \mathbf{\hat{m}_\mathrm{sys}} + \mathbf{\hat{m}_\mathrm{poly}} + \mathbf{\hat{m}_\mathrm{offset}}
\end{equation}
where $\mathbf{\hat{m}_\mathrm{sys}}$ is the systematics component, $\mathbf{\hat{m}_\mathrm{poly}}$ is the polynomial component, and $\mathbf{\hat{m}_\mathrm{offset}}$ is simply the offset term. They are each given by the expressions
\begin{equation} \label{eqn:systematics_model}
\mathbf{\hat{m}_\mathrm{sys}} = \mathbf{X_\mathrm{sys}} 
\cdot 
[w_1\; w_2\; \cdots w_{L}]^\top,
\end{equation}
\begin{equation} \label{eqn:longterm_model}
    \mathbf{\hat{m}_\mathrm{poly}} = \mathbf{X_\mathrm{poly}}
    \cdot
    [w_{L+1}\; w_{L+2}\; \cdots w_{L+P}]^\top,
\end{equation}
\begin{equation} \label{eqn:offset}
    \mathbf{\hat{m}_\mathrm{offset}} = 
    w_{L+P+1}[1\; 1\; \dots 1]^\top.
\end{equation}
Depending on the source, the user can choose which of these components to remove. If the user is interested in preserving long-term trends, such as when obtaining supernova light curves, the de-trended pixel light curve is calculated as $\mathbf{\hat{y}_\mathrm{sig}} = \mathbf{y} - \mathbf{\hat{m}_\mathrm{sys}}$. The user can also subtract the long-term trend $\mathbf{\hat{m}_\mathrm{poly}}$ if, for example, they are interested in transit signals for stars that show long-term variability. If the polynomial component is \textsl{not} used, the design matrix $\mathbf{X} = [\mathbf{X_\mathrm{sys}}\; \mathbf{1}]$ will be of size $N \times (L+1)$, and $\mathbf{w}$ will accordingly be an $(L+1) \times 1$ column vector. The de-trended light curve would also be $\mathbf{\hat{y}_\mathrm{sig}} = \mathbf{y} - \mathbf{\hat{m}_\mathrm{sys}}$. In both cases, subtracting the intercept term $\mathbf{\hat{m}_\mathrm{offset}}$ centers the de-trended light curve around the average flux value.}

\subsection{Ridge Regression} \label{sec:ridge_regression}
As flexible linear models have a tendency to overfit, naively using the OLS method discussed above can result in the systematics model capturing most of the variations in the pixel being modeled, resulting in a de-trended pixel light curve that can resemble white noise. Overfitting can occur as the OLS method attempts to simply find the set of coefficients $\mathbf{\hat{w}}$ that minimizes the discrepancy between the model and data. We note that overfitting does not require that the predictors contain information about the (astrophysical) signal being fitted, and it thus also occurs in our setting where the predictors do not contain any information about the astrophysical signal of the target pixel. We therefore \textsl{regularize} our model to prevent overfitting.

\textsl{Regularization} allows us to prevent overfitting by adding an additional term, called the regularization or penalty term, to an objective function. In \texttt{unpopular} we use \textsl{ridge regression} \citep{hoerl1970}, which involves penalizing large (positive or negative) regression coefficients $w_i$ by adding a regularization term proportional to $\lVert \mathbf{w} \rVert_2^2=\sum_i w_i^2$, the square of the Euclidean ($L^2$) norm of $\mathbf{w}$, to the objective function $S(\mathbf{w})$. This penalization has the effect of shrinking the regression coefficients towards zero.
Ridge regression is effective in alleviating issues related to having correlated regressors in a linear model \citep{hastie2001}.
The systematics model in \texttt{unpopular} contains many correlated regressors since pixels on a given TESS CCD share systematic trends. In addition, as shown below, ridge regression provides a closed-form expression for the ridge coefficients $\mathbf{\hat{w}_{\mathrm{ridge}}}$. Having a closed-form expression is attractive as it removes the need for iterative optimization algorithms and makes the method computationally efficient.

In the simplest case where all coefficients are equally penalized, the regularization term is $\lambda \lVert \mathbf{w} \rVert_2^2$ where $\lambda$ is a non-negative scalar that controls the penalization. A larger value of $\lambda$ results in stronger penalization (i.e., smaller regression coefficients). To equally penalize all the regressors, ridge regression is normally performed on normalized inputs \citep{hastie2001}. We normalize both the regressand and the regressors by dividing by the median and subtracting 1.
After de-trending, this process is inverted to return to the original units. For clarity, the normalized variables will be denoted with an asterisk ($*$) subscript (e.g., $\mathbf{y_*^{}}, \mathbf{X_*^{}}$). After adding the regularization term, the modified objective function $S_{\mathrm{ridge}}(\mathbf{w})$ becomes
\begin{equation}
    S_{\mathrm{ridge}}(\mathbf{w}) = \mathbf{\lVert y_*^{} - X_*^{} \cdot w \rVert_\mathrm{2}^\mathrm{2} + \mathrm{\lambda}\lVert w \rVert_\mathrm{2}^\mathrm{2}}
\end{equation}
and the ridge coefficients $\mathbf{\hat{w}_\mathrm{ridge}}$ are obtained by minimizing $S_{\mathrm{ridge}}(\mathbf{w})$. The formula to calculate $\mathbf{\hat{w}_\mathrm{ridge}}$ is 
\begin{equation} \label{eqn:ridge_simple}
    \mathbf{\hat{w}_\mathrm{ridge}} = (\mathbf{X_*^\top \cdot X_*^{}} + \lambda \mathbf{I})^{-1} \cdot (\mathbf{X_*^\top \cdot y_*^{}}),
\end{equation}
where $\mathbf{I}$ is an identity matrix with the same dimensions as $(\mathbf{X_*^\top \cdot X_*^{}})$. \edit1{When $\lambda \rightarrow 0$ (i.e., no penalization), the coefficients $\mathbf{\hat{w}_\mathrm{ridge}}$ are equivalent to the OLS solution.}

We can generalize the above formulation to allow for separate specification of the penalization for each coefficient $w_i$. In other words, there can be a (possibly) different $\lambda_i$ for each $w_i$. For concreteness, we will assume the polynomial component is being used and that the size of $\mathbf{w}$ is \edit1{$L+P+1$}. The generalization involves replacing the previous regularization parameter $\lambda$ with a matrix. We will denote this matrix as $\mathbf{\Gamma}$ and it will be an \edit1{$(L+P+1) \times (L+P+1)$} square and diagonal matrix   
\begin{equation}
    \mathbf{\Gamma} = 
    \begin{bmatrix}
    \gamma_1 & & \\ 
    & \ddots & \\
    & & \gamma_{\scriptscriptstyle{L+P+1}} \\ 
    \end{bmatrix},
\end{equation}
where each element $\gamma_i$ will turn out to be the square root of the regularization parameter $\lambda_i$. The term added to $S(\mathbf{w})$ \edit1{is now} $\mathbf{\lVert \Gamma \cdot w \rVert_\mathrm{2}^\mathrm{2}}$, and the modified objective function $S_{\mathrm{ridge}}(\mathbf{w})$ is
\begin{align}
    S_{\mathrm{ridge}}(\mathbf{w}) = \mathbf{\lVert y_*^{} - X_*^{} \cdot w \rVert_\mathrm{2}^\mathrm{2} + \lVert \Gamma \cdot w \rVert_\mathrm{2}^\mathrm{2}}.  
\end{align}
As shown in Appendix \ref{appendix:ridge_regression}, the coefficients $\mathbf{\hat{w}_\mathrm{ridge}}$ can be calculated with the formula
\begin{equation}
    \mathbf{\hat{w}_\mathrm{ridge}} = (\mathbf{X_*^\top \cdot X_*^{}} + \mathbf{\Gamma^\top \cdot \Gamma})^{-1} \cdot (\mathbf{X_*^\top \cdot y_*^{}}).
\end{equation}
We can rewrite this formula to more closely parallel Equation (\ref{eqn:ridge_simple}) by defining $\mathbf{\Lambda \equiv \Gamma^\top \cdot \Gamma}$ such that 
\begin{equation}
    \mathbf{\Lambda} = 
    \begin{bmatrix}
    \lambda_1 & & \\ 
    & \ddots & \\
    & & \lambda_{\scriptscriptstyle{L+P+1}} \\ 
    \end{bmatrix},
\end{equation}
where each element $\lambda_i = \gamma_i^2$. The formula for the coefficients $\mathbf{\hat{w}_{\mathrm{ridge}}}$ is then
\begin{equation}
    \mathbf{\hat{w}_\mathrm{ridge}} = (\mathbf{X_*^\top \cdot X_*^{}} + \mathbf{\Lambda})^{-1} \cdot (\mathbf{X_*^\top \cdot y_*^{}}).
\end{equation}
The simplest case of equal penalization is when $\lambda=\lambda_i$ and $\mathbf{\Lambda} = \lambda \mathbf{I}$. \edit1{Larger values of $\lambda_i$ lead to stronger regularization (i.e., smaller $\lvert w_i \rvert$), and can hence prevent overfitting.}

In \texttt{unpopular}, we can currently specify up to two regularization parameters $\lambda_\mathbf{L}$ and $\lambda_\mathbf{P}$ depending on whether the coefficient is for a predictor pixel light curve (i.e., the systematics model) or for a polynomial component (i.e., astrophysical long-term trend). All the predictor pixel light curve regressors will be penalized with $\lambda_\mathbf{L}$, while the polynomial components will be penalized with $\lambda_\mathbf{P}$. The intercept term, \edit1{the column of 1s}, is left unpenalized \citep{hastie2001}.
The full \edit1{$(L+P+1) \times (L+P+1)$} regularization matrix $\mathbf{\Lambda}$ can be constructed as 
\begin{equation}
    \mathbf{\Lambda} = 
    \begin{bmatrix}
    \mathbf{\Lambda_\mathbf{L}} & & \\ 
    & \mathbf{\Lambda_\mathbf{P}} & \\
    & & 0
    \end{bmatrix},
\end{equation}
where $\mathbf{\Lambda_\mathbf{L}}$ is an $L \times L$ square and diagonal matrix with all diagonal elements set to $\lambda_\mathbf{L}$ ($\mathbf{\Lambda_L = \lambda_L I}$), and $\mathbf{\Lambda_\mathbf{P}}$ is a \edit1{$P \times P$} square and diagonal matrix with all diagonal elements set to $\lambda_\mathbf{P}$ ($\mathbf{\Lambda_P = \lambda_P I}$). The final 0 value indicates that the intercept term is not regularized. 

We now have a different set of coefficients $\mathbf{\hat{w}_\mathrm{ridge}}$ from using ridge regression instead of OLS. The rest of the approach is the same. The model is $\mathbf{\hat{m} = X_*^{} \cdot \hat{w}_\mathrm{ridge}}$, and if the polynomial component was used, $\mathbf{\hat{m}}$ can be split up into the systematics component $\mathbf{\hat{m}_\mathrm{sys}}$, the astrophysical long-term component $\mathbf{\hat{m}_\mathrm{poly}}$, and the intercept term $\mathbf{\hat{m}_\mathrm{offset}}$ using Equations (\ref{eqn:systematics_model}), (\ref{eqn:longterm_model}), and (\ref{eqn:offset}). Similar to before, after subtracting $\mathbf{\hat{m}_\mathrm{sys}}$ from $\mathbf{y}_*^{}$ to remove the systematics, the user can also subtract $\mathbf{\hat{m}_\mathrm{poly}}$ to remove any captured long-term trend and subtract $\mathbf{\hat{m}_\mathrm{offset}}$ to center the de-trended light curve around its average value. 

\subsection{Train-and-Test Framework} \label{sec:train_and_test_framework}
In \texttt{unpopular} we employ a modified version of the ``train-and-test"\footnote{While we use this terminology to indicate that the dataset our model is trained on and the dataset being de-trended are disjoint sets, we do \textsl{not} calculate a statistic for each test set to assess the model performance as is commonly done in train-and-test frameworks.} framework used in the original \textsl{CPM} method \citep{wang2016} to further prevent overfitting.

\begin{figure}
    \centering
    \includegraphics[width=\textwidth]{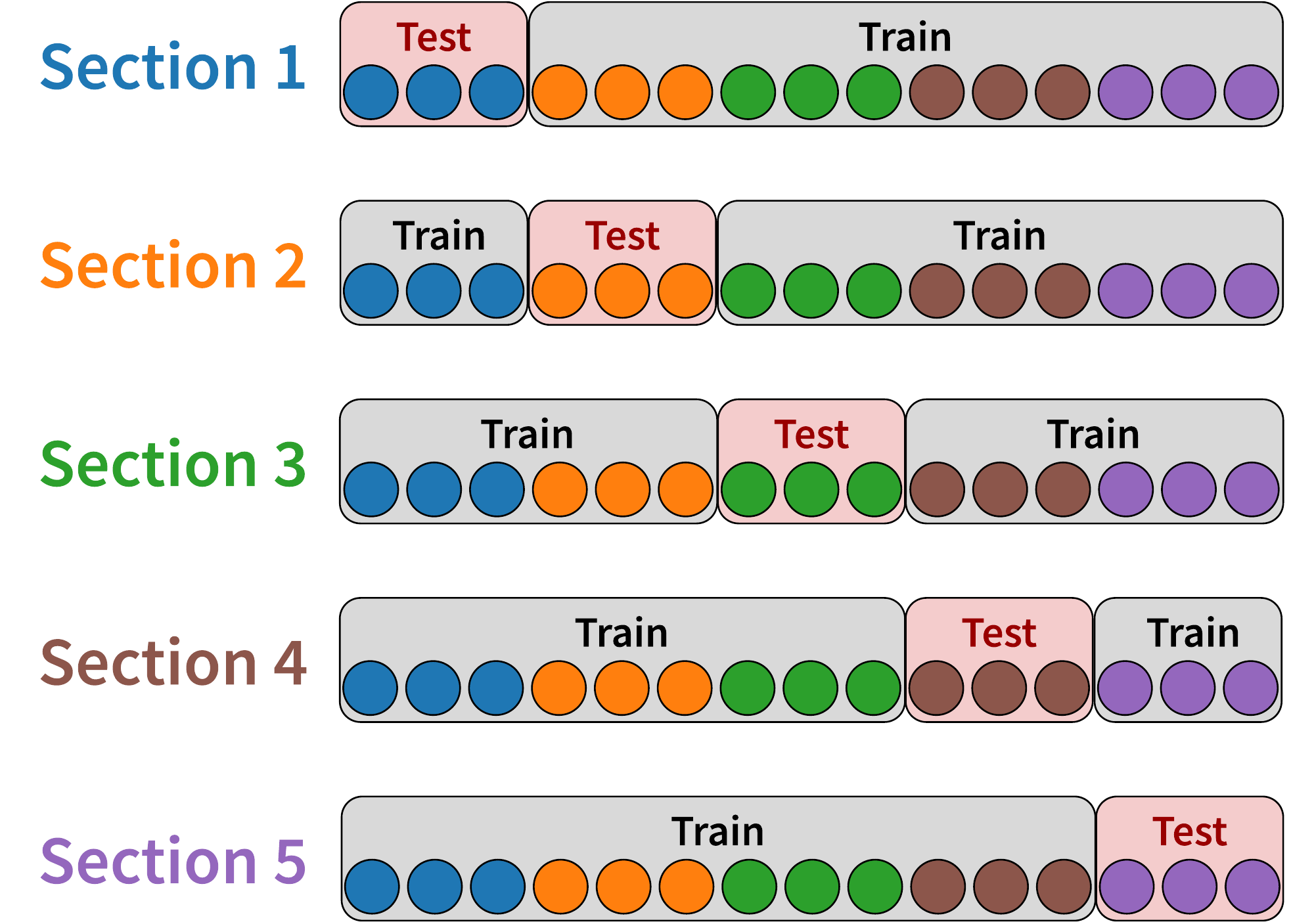}
    \caption{A diagram of the train-and-test framework. In this scenario, the light curve being de-trended is divided into five contiguous sections ($k=5$), each indicated by a separate color. For every test section being de-trended the model coefficients are obtained by fitting to the training data, which is the rest of the light curve. As $k$ is inversely proportional to the number of measurements in a given section, an appropriate choice for $k$ can be made is based on the timescale of the signal of interest.}
    \label{fig:train_test_framework}
\end{figure}

Under this framework, the target pixel data used to calculate the coefficients $\mathbf{\hat{w}_\mathrm{ridge}}$, the \textsl{training} data, are \edit1{temporally separated} from the target pixel data being predicted and de-trended, the \textsl{test} data. As the coefficients are calculated using the training data and the de-trending occurs on the test data, the model \edit1{is less likely to} overfit signals with timescales shorter than the test data window. The separation is accomplished by splitting the target pixel light curve into $k$ contiguous sections \edit1{and by calculating a different $\mathbf{\hat{w}_\mathrm{ridge}}$ for each section (see Figure \ref{fig:train_test_framework})}.

\edit1{In the original \textsl{CPM} method $k{\sim}4350$ as it predicted each flux measurement from a single-quarter (three months) Kepler light curve independently, and \citet{wang2016} estimated that de-trending would take take 7 hours on a single core machine. In addition, several flux measurements before and after (in time) the flux measurement being predicted were excluded to decrease the chance of overfitting to the potential transit signal. These choices were made to increase the sensitivity to faint transits that could be potential Earth analogs. As the detection of faint transits is not a priority for \texttt{unpopular} we do not require this level of sensitivity. We therefore set $k{\sim}100{-}200$ for a single-sector (${\sim}27$ day) TESS light curve to significantly increase the efficiency and do not exclude any neighboring flux measurements. We allow the user to specify the number of sections and discuss the effect of changing this tuning parameter in \S\ref{sec:tuning_parameters}.}

\section{Creating De-trended TESS light curves with \texttt{unpopular}} \label{sec:unpopular}

In this section we describe the steps for using \texttt{unpopular} to extract TESS FFI light curves. We will use the Sector 1 data for ASASSN-18tb/SN 2018fhw \citep{vallely2019}, a Type-Ia supernova event observed by TESS, as an example. The coordinates for this source are RA, Dec (J2000): $64.525833^\circ$, $-63.615669^\circ$.\footnote{\url{https://www.wis-tns.org/object/2018fhw}}
While there are several tuning parameters (i.e., values and choices the user must make \textsl{before} de-trending) in these steps, we simply note the tuning parameters used for ASASSN-18tb during each of the steps and postpone discussing the effects of changing these parameters to \S\ref{sec:tuning_parameters}. 

\subsection{Downloading Data} \label{sec:data_download}

TESS obtains time-series photometry with its 16 CCDs (four CCDs per camera) in a given observing sector. Each CCD has $2048 \times 2048$ science pixels, and the full FOV was recorded at either 30-minute cadence (Cycles 1 \& 2) or 10-minute cadence (Cycle 3). The stack of all FFIs for a single CCD in a given sector taken with 30-minute cadence requires ${\sim}45$ GB of storage. As files of this size are not user-friendly and since \texttt{unpopular} only requires a region of the FFIs around a given source, we recommend downloading a cutout stack instead of a full FFI stack (see Figure \ref{fig:scattered_light_asassn18tb}). Calibrated FFI cutout stacks of a specified size centered at a given coordinate can be downloaded using tools such as \texttt{TESScut} \citep{brasseur2019} or \texttt{lightkurve} \citep{barentsen2020}. The cutouts must be large enough to allow for enough predictor pixels to be chosen even when taking into account that a set of pixels close to the target pixel will be excluded. Based on our experience, we find that $100 \times 100$ FFI cutouts (${\sim}250$ MB for a stack of 30-minute cadence FFIs) are reasonable for sources with aperture regions smaller than $5 \times 5$ pixels. 
While we have not performed any tests, we do not believe our package is appropriate for obtaining light curves from saturated targets that illuminate a large region of the FFI cutouts. For users interested in saturated targets, the ``halo-photometry" approach taken by \citet{white2017} and \citet{pope2019} will likely be more successful.

\subsection{Pre-processing} \label{sec:pre-processing}
The TESS team provides \texttt{QUALITY} arrays that flag detected anomalies\footnote{\url{https://archive.stsci.edu/missions/tess/doc/EXP-TESS-ARC-ICD-TM-0014-Rev-F.pdf}} (e.g., spacecraft is in coarse point mode, reaction wheel desaturation event) in the data. The default behavior in \texttt{unpopular} is to remove the FFI cutouts with any quality flags prior to de-trending as they tend to be outliers (see Figure \ref{fig:scattered_light_asassn18tb}). The quality flags used here are those recorded in the FFI data, which are applicable to the \textsl{entire} CCD that the source was observed on. \edit1{While additional quality flags were recorded for two-minute cadence targets in the Target Pixel Files (TPFs) and can be mapped to the FFIs, an approach taken by the \texttt{eleanor} pipeline, we do not default to using them.}

After this step, \texttt{unpopular} calculates and stores the normalized version of the pixel light curves for all pixels in the FFI cutout. This step is necessary as, discussed in \S\ref{sec:ridge_regression}, ridge regression requires us to normalize both the regressand and the regressors. We divide each pixel light curve by its median value and then subtract 1
from the resulting light curve to center it around zero. This normalization makes the flux measurements unitless and we invert this process after de-trending to recover the physical units. We also allow the user to perform an initial background subtraction before de-trending, where we first obtain the median light curve of the 100 faintest pixels in the FFI cutout then subsequently subtract it from the entire cutout. While this initial background subtraction does not affect the shape of the de-trended light curve, it may rescale the light curve to a more accurate baseline flux level after de-trending.
However, as we mention below the baselines fluxes are rough estimates and should not be treated as accurate. We did \textsl{not} perform the initial background subtraction for this example. 

\subsection{Model Specification} \label{sec:model_specification}

This step involves specifying the aperture, constructing the design matrix, and setting the regularization parameters.

\subsubsection{Aperture Specification} \label{sec:aperture_selection}
\begin{figure}
    \centering
    \includegraphics[width=\textwidth]{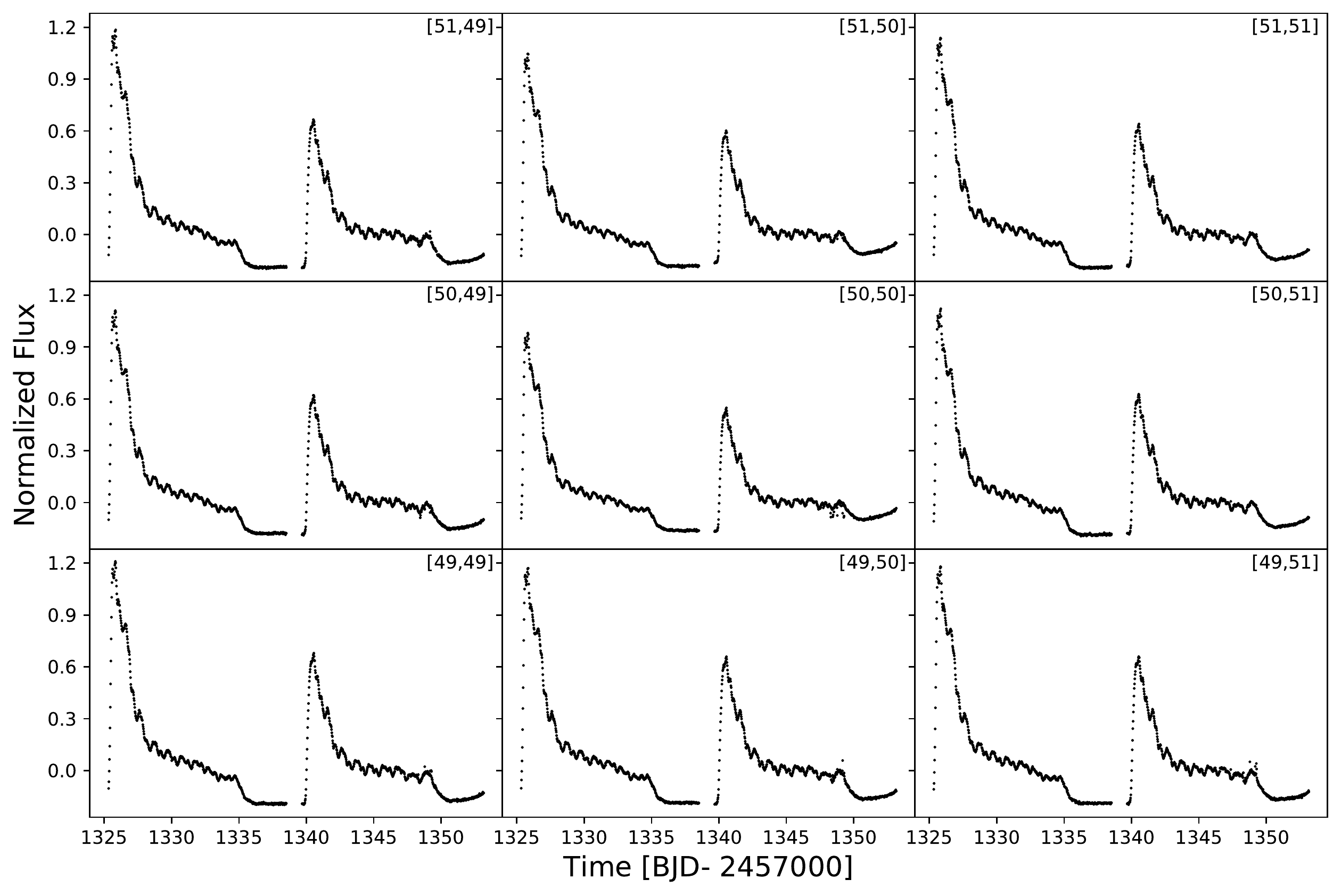}
    \caption{The normalized light curve for each pixel in the $3 \times 3$ aperture of ASASSN-18tb. The numbers in the top right of each panel indicate the pixel row and column positions. 
    }
    \label{fig:pixbypix_normflux_asassn18tb}
\end{figure}

While the method explained in \S\ref{sec:cpm} de-trends a single pixel light curve, for most scientific studies the user will want an aperture light curve. As \texttt{unpopular} produces a de-trended FFI cutout of a chosen rectangular set of pixels, the aperture light curve can be obtained by summing all the de-trended pixels or a subset of them. For every pixel being de-trended a \textsl{separate} linear model, each with its own set of predictor pixels and coefficients, is constructed. For ASASSN-18tb, we will use a $3 \times 3$ square aperture covering rows 49 to 51 and columns 49 to 51, or in detector coordinates, rows 802 to 804 and columns 730 to 732 (Figure \ref{fig:pixbypix_normflux_asassn18tb}). The location of the source falls on pixel [50,50]. The $3 \times 3$ aperture was chosen by visually inspecting the region around the source in the FFI cutout and by choosing an aperture that encompassed a set of contiguous pixels near the central pixel that were brighter than surrounding empty regions. While this approach is not quantitative nor easily programmable, developing a robust quantitative approach to automate aperture selection is beyond the scope of this work.

\subsubsection{Constructing the Design Matrix} \label{sec:design_matrix_construction}
Constructing the design matrix consists of selecting the predictor pixels and deciding on whether to include or exclude the polynomial component. As predictor pixels must not be from the same source, we first set an exclusion region around each pixel where no predictor pixels are chosen from. The default exclusion region is an $11 \times 11$ square set of pixels centered on each target pixel \edit1{, corresponding to a ${\sim}3.85\arcmin \times {\sim}3.85\arcmin$ region given the $21\arcsec$ TESS pixels. We made this choice as we do not expect most unsaturated choices to illuminate a larger region.}. For sources that do, the user will need to increase the size of the exclusion region. 

\begin{figure}
    \centering
    \includegraphics[width=0.495\textwidth]{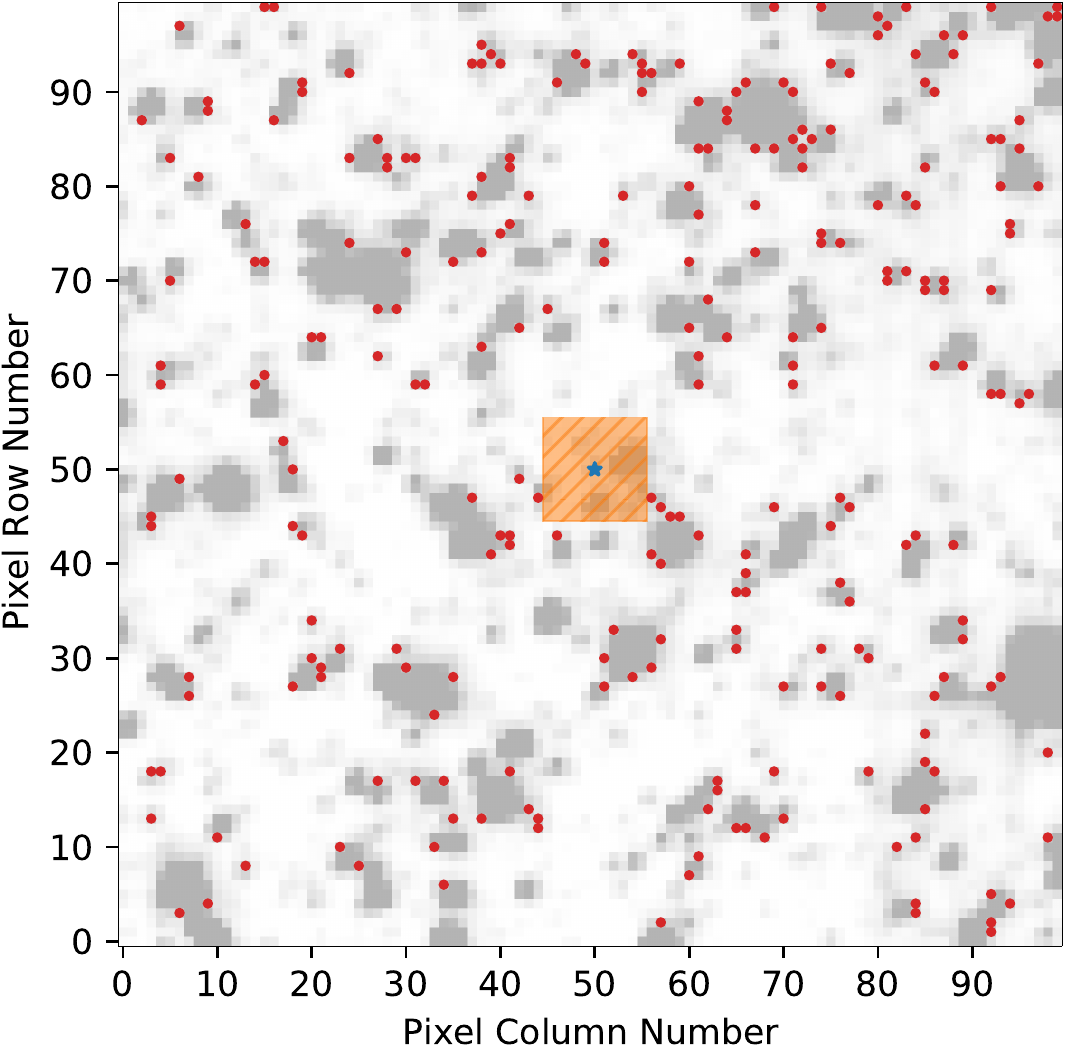}
    \includegraphics[width=0.495\textwidth]{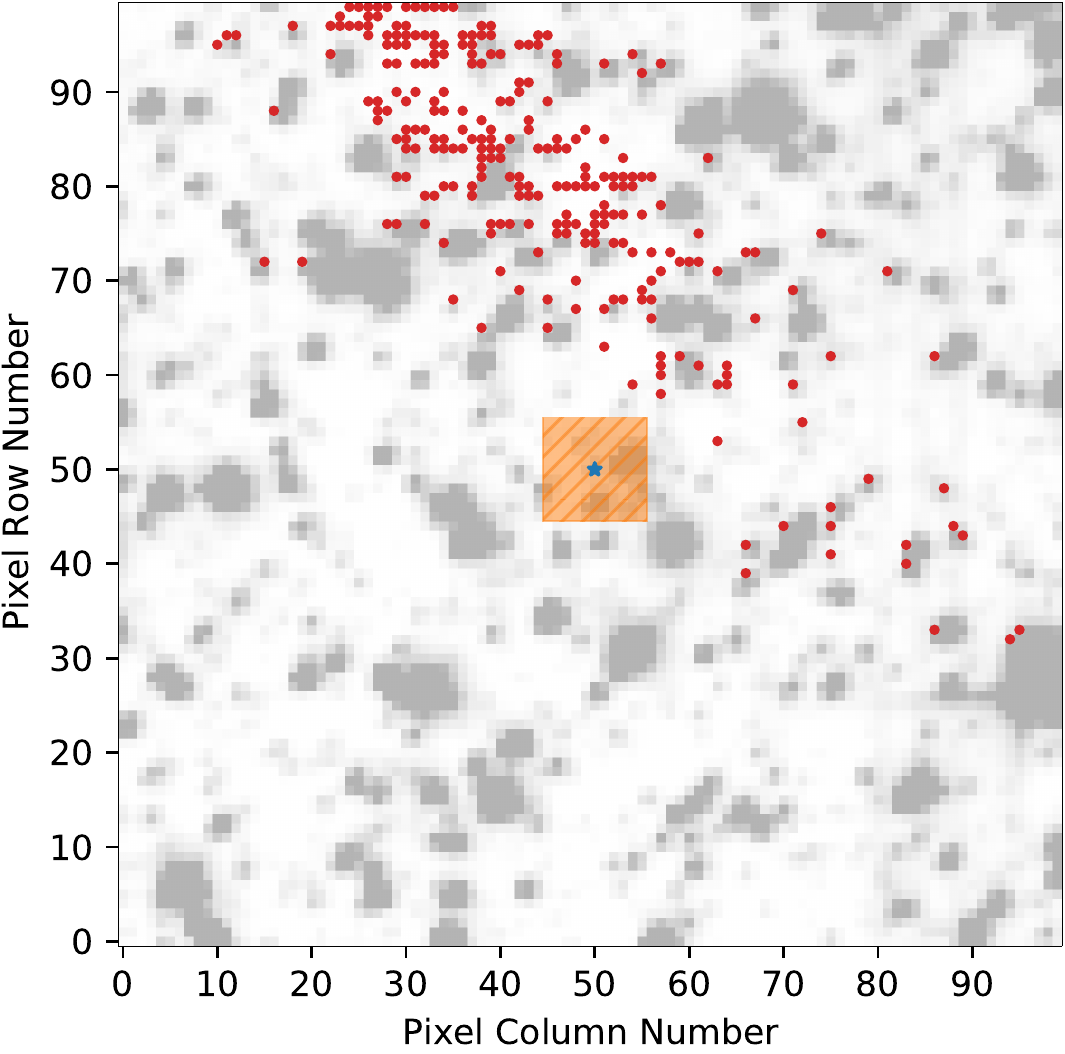}
    \caption{Difference between the choice of predictor pixels based on the selection method. In both panels, the pixel being de-trended is the central pixel (blue star).
    The exclusion region around the central pixel (shaded and hatched orange area) is where predictor pixels are \textit{not} chosen from. In the left panel, the predictor pixels (red circles) were chosen based on having a similar median brightness to the pixel being de-trended. In the right panel, the predictor pixels were chosen based on having a high cosine similarity to the pixel being de-trended. The clustering of predictor pixels in the right panel indicates there are local systematic effects in this region.}
    \label{fig:predictor_pixels}
\end{figure}

Once the exclusion region is set we must decide on the number of predictor pixels $L$ and the method to choose them. While these decisions will be the same across all the pixels in the aperture, we reiterate that the set of chosen predictor pixels will be different for each target pixel. The number of predictor pixels must be large enough to allow for a flexible model that can capture most of the systematic effects. After experimentation, we found that the de-trending results were mostly similar for values of $L$ between 64--256. Setting $L$ below this range resulted in the de-trended light curve \edit1{still showing the two large scattered light effects} as the model was inflexible. Setting $L$ above this range increased the runtime (i.e., computational time) while showing similar de-trending results. \edit1{Based on these qualitative observations we} chose $L=256$ predictor pixels as the default setting and find this value to be acceptable for many sources. 

In Figure \ref{fig:predictor_pixels} we show two of the currently implemented methods for selecting the predictor pixels. In the left figure, the predictor pixels are the top $L=256$ pixels with the closest median brightness to the target pixel. This method is based on the idea that \edit1{photon noise properties and} systematic effects may be more similar across pixels of similar brightness \edit1{potentially due to CCD (non)-linearity\footnote{\url{https://archive.stsci.edu/files/live/sites/mast/files/home/missions-and-data/active-missions/tess/_documents/TESS_Instrument_Handbook_v0.1.pdf} (6.4.2)}} and is the default method. 
In the right figure, the predictor pixels are the top $L=256$ pixels with the highest cosine similarity (i.e., similar trend) to the target pixel. The fact that the predictor pixels are clustered in one area indicates that there are local systematic effects in this region. These local systematic effects, likely caused by the change in proximity of the spacecraft to Earth during an orbit, tend to ``sweep" from one side of the FOV to the other.\footnote{For examples of these systematic effects, we recommend viewing the TESS: The Movies videos on \url{https://www.youtube.com/c/EthanKruse/videos}} Although these ``sweeping" signals create an opportunity to use time-lagged versions of pixel light curves as regressors, we postpone exploring this approach to a future study. While the set of predictor pixels are significantly different between these two methods, we found that the de-trending results were similar regardless of which method was chosen. We also found that randomly choosing the predictor pixels worked as well as the other methods. For ASASSN-18tb, we used the default settings for both the number and selection method of the predictor pixels.

The final step in constructing the design matrix is choosing whether to include or exclude the polynomial component. A single polynomial is used for the whole Sector of data. When the source is expected to show long-term astrophysical trends, such as a supernovae or other explosive transients,
including the polynomial component mitigates distorting the de-trended light curve.
Erroneously excluding the polynomial component when there is long-term astrophysical variability will likely result in a reduction of the amplitude of the long-term signal and distortions in other parts of the light curve as the systematics model will attempt to fit to the astrophysical signal. If the user chooses to include a polynomial component, the default setting creates a cubic polynomial \edit1{simply because it is the lowest degree polynomial that is non-linear and allows for asymmetry}. As the polynomial is simply there to capture long-term astrophysical trends, in general we do not recommend the use of more flexible higher degree polynomials. For this example we included the default cubic polynomial.

\subsubsection{Setting the Regularization Parameters}
After the design matrix has been constructed, the regularization parameters must be set. The regularization parameters are set separately for the systematics model $\lambda_\mathbf{L}$ and the polynomial model $\lambda_\mathbf{P}$ (if using). As mentioned in \S\ref{sec:cpm}, the regularization parameters are defined as the \textsl{precision}, the reciprocal of the variance, of the Gaussian prior for each regressor. Therefore, larger values of $\lambda$ lead to stronger regularization. For ASASSN-18tb, we set $\lambda_\mathbf{L}=0.01$ and $\lambda_\mathbf{P}=0.01$. The effects of changing these values are discussed in \S\ref{sec:tuning_parameters}.

\subsection{De-trending} \label{sec:de-trending}
\begin{figure}
    \centering
    \includegraphics[width=\textwidth]{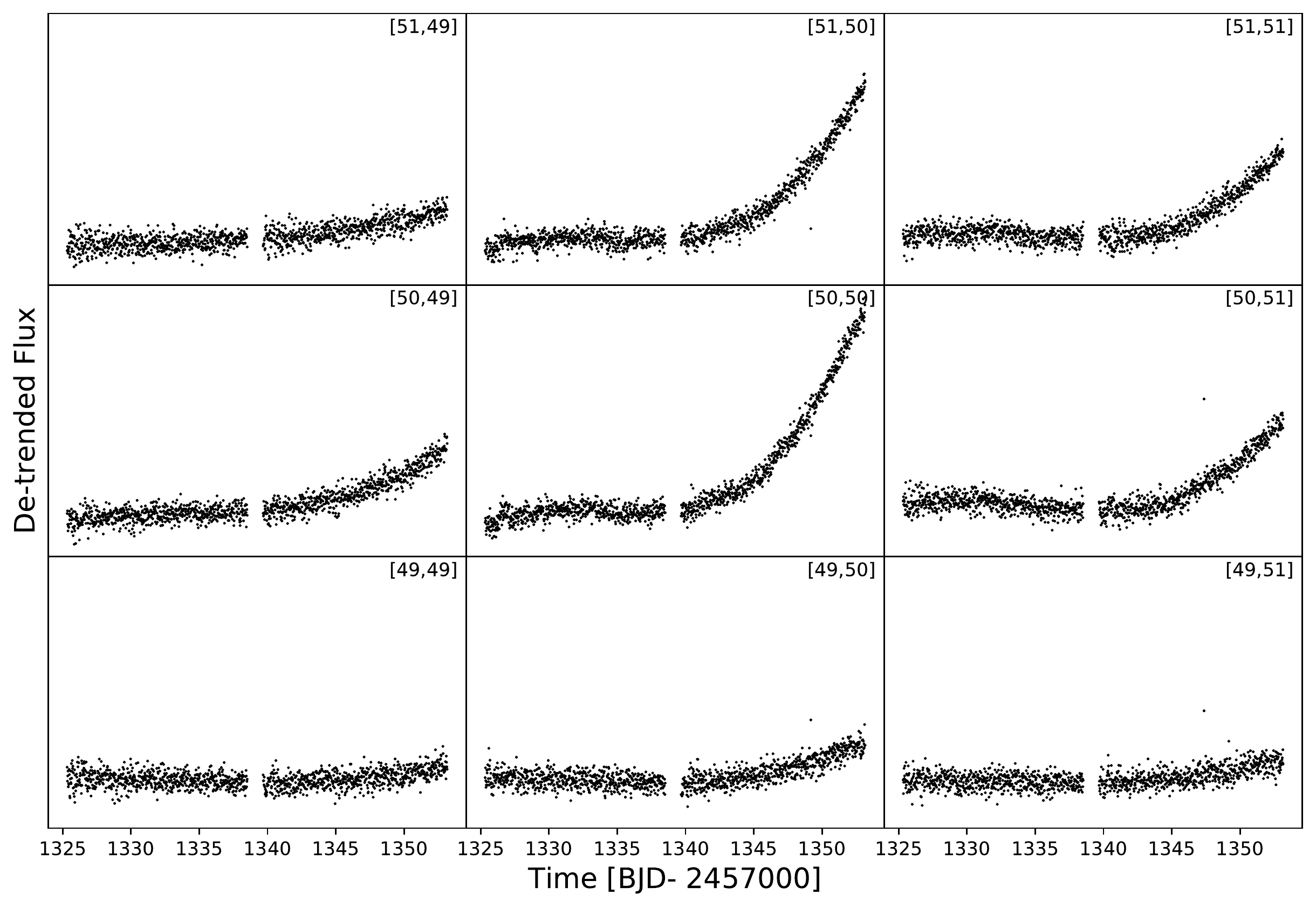}
    \includegraphics[width=\textwidth]{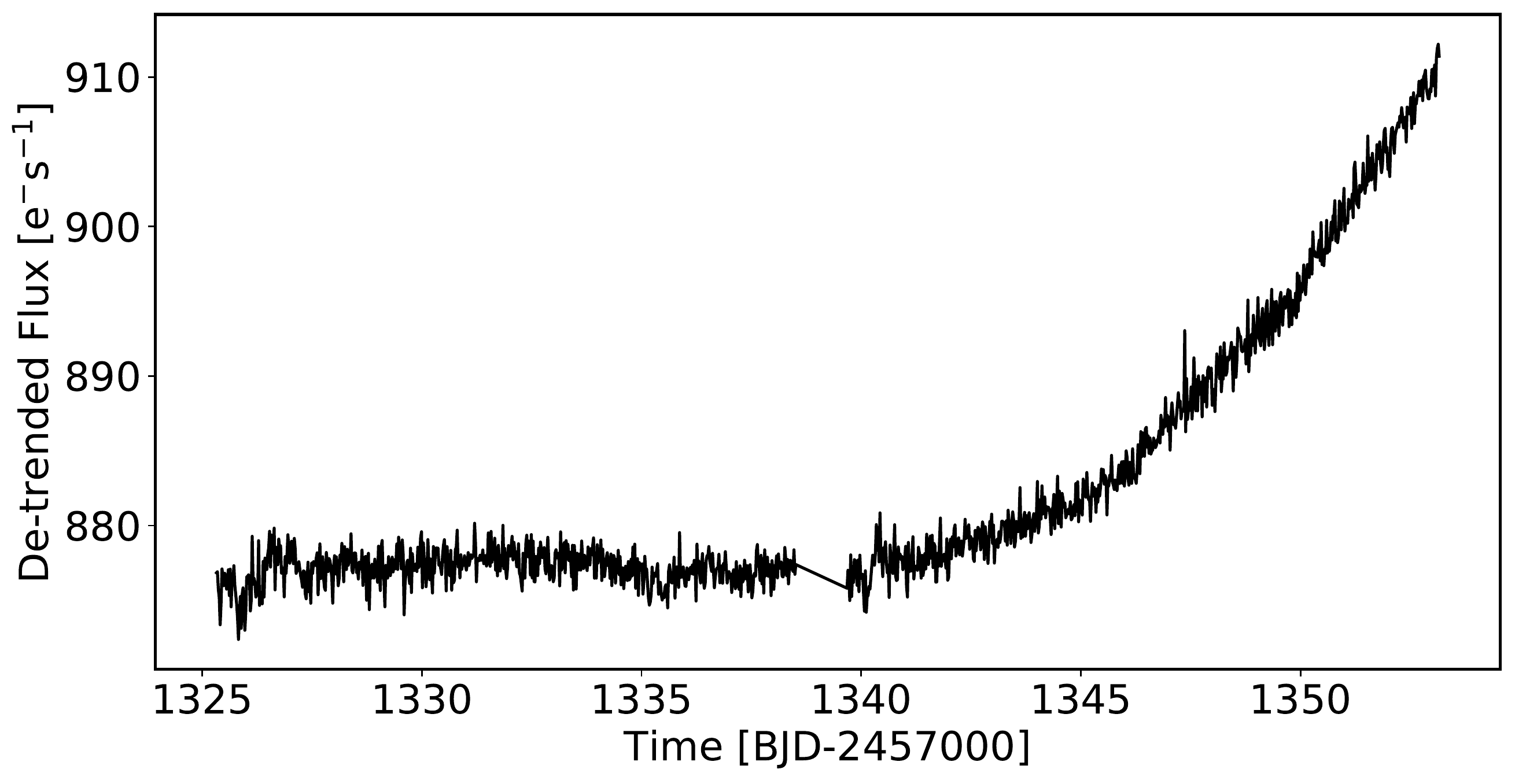}
    \caption{The de-trended pixel light curves (top panel) and the de-trended aperture light curve (bottom panel). Each of the light curves in the top panel are the de-trended versions of those shown in Figure \ref{fig:pixbypix_normflux_asassn18tb}. 
    The aperture light curve is obtained by first scaling each pixel light curve to its estimated physical flux and then summing them together. No initial background subtraction was performed, a cubic polynomial was added to capture the long-term trend, and we set $L=256, \lambda_\mathbf{L}=0.01, \lambda_\mathbf{P}=0.01, k=126$.}
    \label{fig:pixbypix_detrended_asassn18tb}
\end{figure}

The final tuning parameter to set before de-trending is $k$, the number of sections to use for the train-and-test framework (see \S\ref{sec:train_and_test_framework}). The smallest possible choice is $k=2$, which separates the light curves into two sections. The largest possible choice is when $k$ is equal to the number of data points, where a separate set of coefficients $\mathbf{\hat{w}_\mathrm{ridge}}$ is calculated to predict each data point. We set $k=126$ (approximately ten data points for each section) for ASASSN-18tb. We de-trend each pixel inside the aperture by fitting the linear model defined in \S\ref{sec:cpm} and subtracting the systematics component. The de-trending took ${\sim}5$ seconds on a Dell XPS 15 9570 laptop (Intel Core i7-8750H 2.20 GHz) and we show the results in Figure \ref{fig:pixbypix_detrended_asassn18tb}. The top plot shows each of the de-trended pixel light curves and the bottom plot shows the aperture light curve. We obtain the aperture light curve by summing all the rescaled de-trended pixel light curves. The rescaling is done by inverting the normalization, or in other words by multiplying each de-trended pixel light curve by its original median value and then adding back the median value. We caution the user that we do not expect adding the median flux will shift the de-trended light curve to the ``true" flux level. Shifting the de-trended light curve to an accurate flux level will likely require comparing the de-trended light curve to a known comparison object in the FOV (i.e., relative photometry) and is beyond the scope of this project. Our aperture de-trended light curve is similar to that obtained by \citet{vallely2019}, where they used the ISIS image subtraction package \citep{alard1998, alard2000}. 

\subsection{Tuning Parameters} \label{sec:tuning_parameters}
We discuss the effects of modifying the tuning parameters (i.e., hyperparameters) here. We restrict the discussion to three tuning parameters: the regularization parameters $\lambda_\mathbf{L}$ and $\lambda_\mathbf{P}$, and the number of sections $k$ for the train-and-test framework. 
We will use ASASSN-18tb's central pixel ($[50, 50]$), the pixel with the largest variation after de-trending (see Figure \ref{fig:pixbypix_detrended_asassn18tb}), to demonstrate the effect of changing the tuning parameters. 

To quantify the effects of changing these tuning parameters, we calculated a proxy for the 6-hr combined differential photometric precision (CDPP) \citep{christiansen2012} on the de-trended pixel light curve. CDPP is a photometric noise metric in units of parts-per-million (ppm) developed by the Kepler team to define the ease of detecting a weak terrestrial transit signature at a given timescale in a light curve. A smaller value is preferred as CDPP is a characterization of the noise. The proxy we use for the 6-hr CDPP, based on the approach by \citet{gilliland2011}, is easier to calculate than the formal wavelet-based CDPP and is implemented in the \texttt{lightkurve} package. The method first fits and subtracts a running 2-day quadratic polynomial (i.e., Savitzky–Golay filter) from the de-trended light curve to remove low-frequency signals. After this step, sigma-clipping at $5\sigma$ is done to remove outliers. The proxy CDPP is then calculated as the standard deviation of the 6-hr (12 measurements with 30-minute cadence data) running mean. We refer to this value as the 6-hr CDPP hereafter. 

While we calculate CDPP, we note that for the example source, which is a supernova, CDPP is \textit{not} a particularly good metric to assess the quality of the de-trended light curve. CDPP was developed for assessing the ease of detecting Earth-like transit signatures and is the appropriate metric to optimize when de-trending light curves for exoplanet searches. For users interested in obtaining supernova light curves, stellar rotation periods, or investigating stellar variability in general, de-trending based on minimizing CDPP risks removing signals of interest. However, as we find that smaller values of CDPP tend to go hand in hand with qualitatively good light curves, we use CDPP for demonstrative purposes. While there exist other metrics developed with studying stellar variability in mind \citep{basri2013} we defer investigating these metrics to a future paper. 

\begin{figure}
    \centering
    \includegraphics[width=\textwidth]{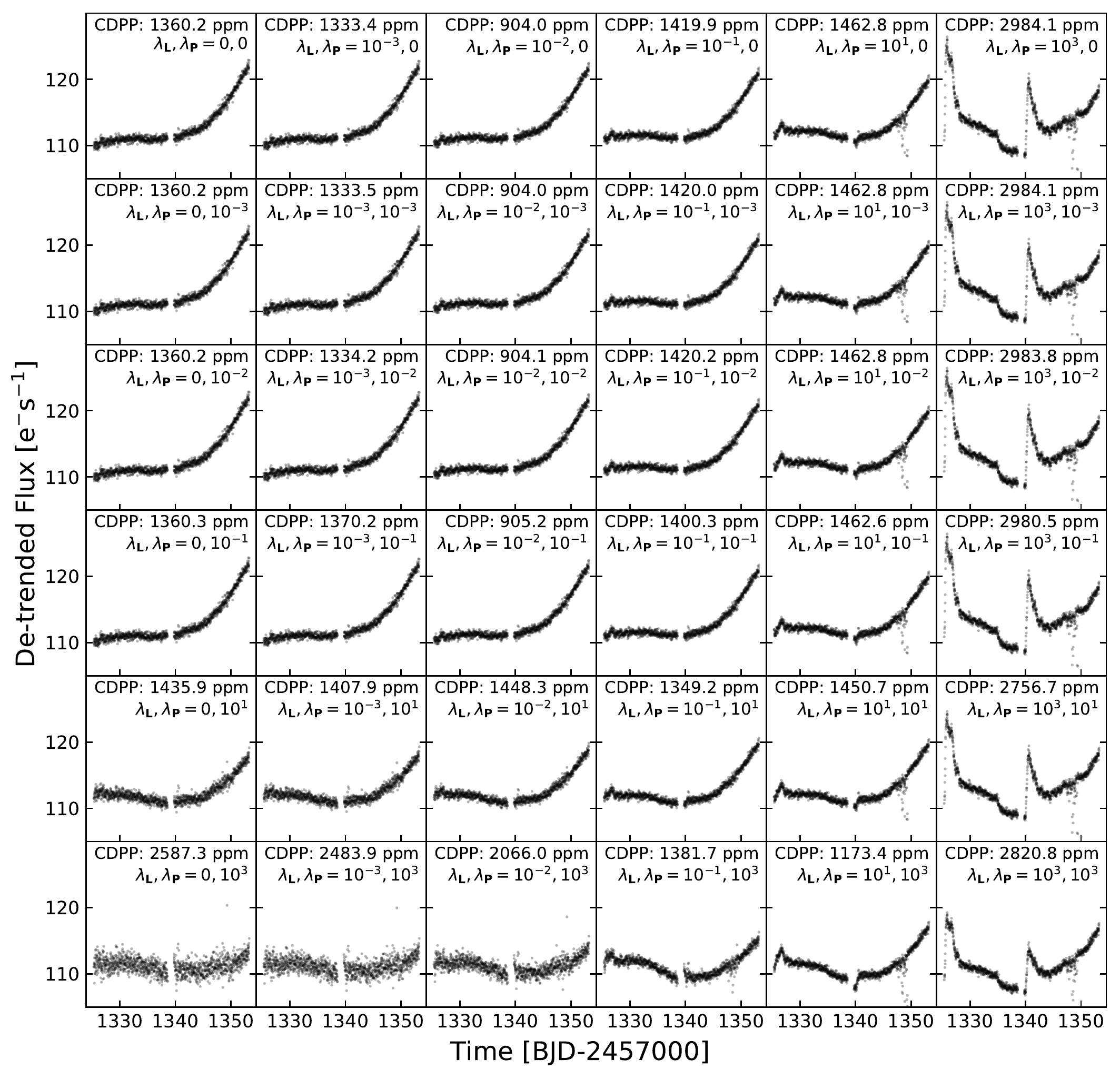}
    \caption{Different de-trending results for the central pixel of ASASSN-18tb when varying $\lambda_\mathbf{L}$ and $\lambda_\mathbf{P}$ while fixing $L=256, k=126$. Each panel shows the de-trended pixel light curve and the 6-hr CDPP for a specific combination of $\lambda_\mathbf{L}$ and $\lambda_\mathbf{P}$. The value of $\lambda_\mathbf{L}$ increases from the left column to the right column, and the value of $\lambda_\mathbf{P}$ increases from the top row to the bottom row. The smallest 6-hr CDPP values of ${\sim}904$ ppm are obtained in the top four rows of the third column where $\lambda_\mathbf{L}=10^{-2}$ and $\lambda_\mathbf{P}=0,10^{-3},10^{-2},10^{-1}$. The bottom-left panels show signs of the systematic component overfitting to the long-term trend, while the right columns show signs of underfitting}
    \label{fig:comparing_reg_parameters}
\end{figure}

\begin{figure}
    \centering
    \includegraphics[width=\textwidth]{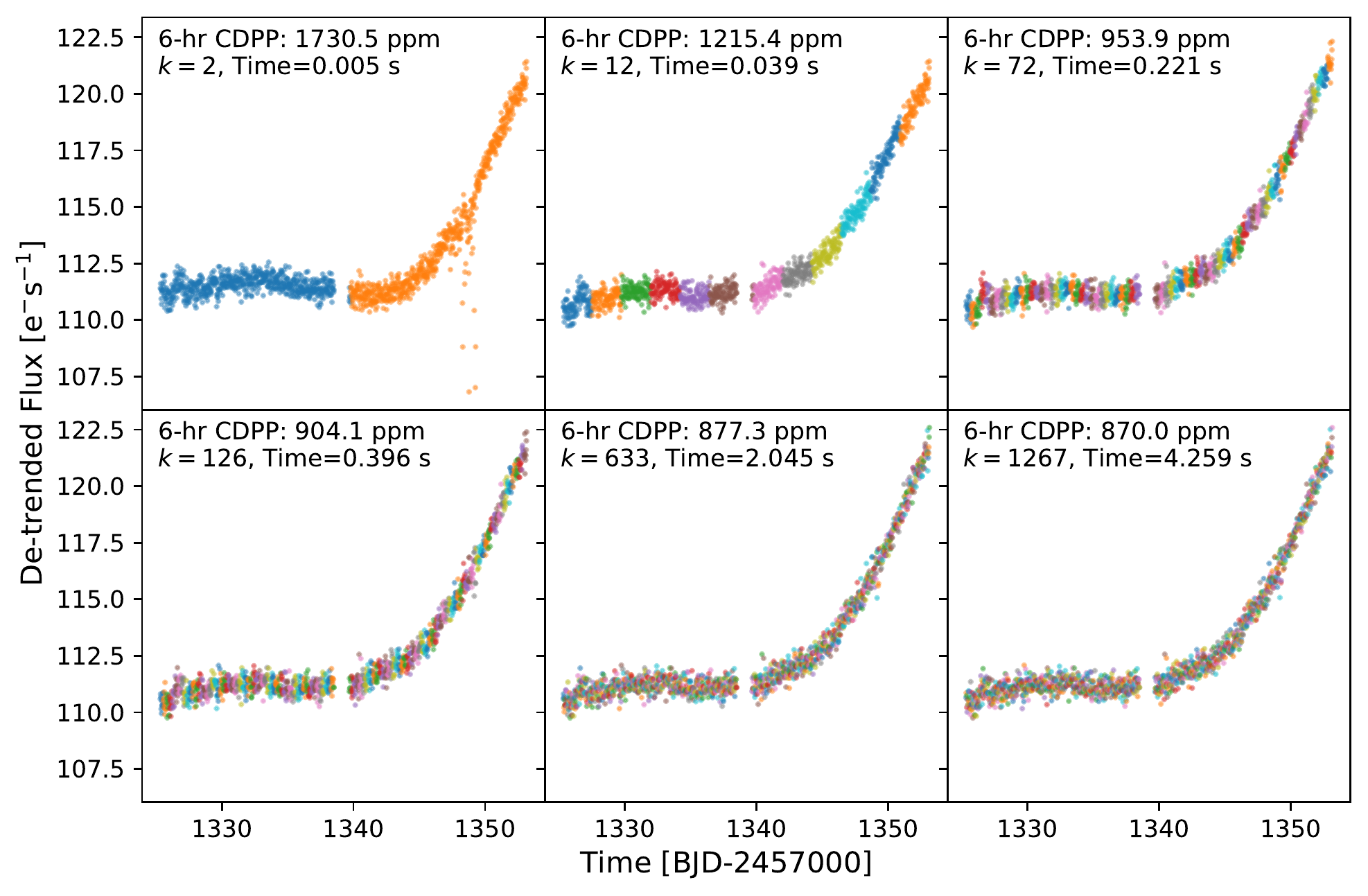}
    \caption{Different de-trending results for the central pixel of ASASSN-18tb when varying $k$, the number of sections to use for the train-and-test framework. Other tuning parameters were kept constant ($L=256, \lambda_\mathbf{L}=\lambda_\mathbf{P}=10^{-2}$). Each section is indicated by \edit1{contiguous} points of the same color (non-contiguous points of the same color are \textsl{different} sections). As $k$ becomes larger, the 6-hr CDPP decreases while the runtime increases. While most values of $k$ show similar results, when $k=2$ the systematic effects around 1348-1350 days are still present in the de-trended light curve.}
    \label{fig:comparing_kvalues}
\end{figure}

\begin{figure}
    \centering
    \includegraphics[width=\textwidth]{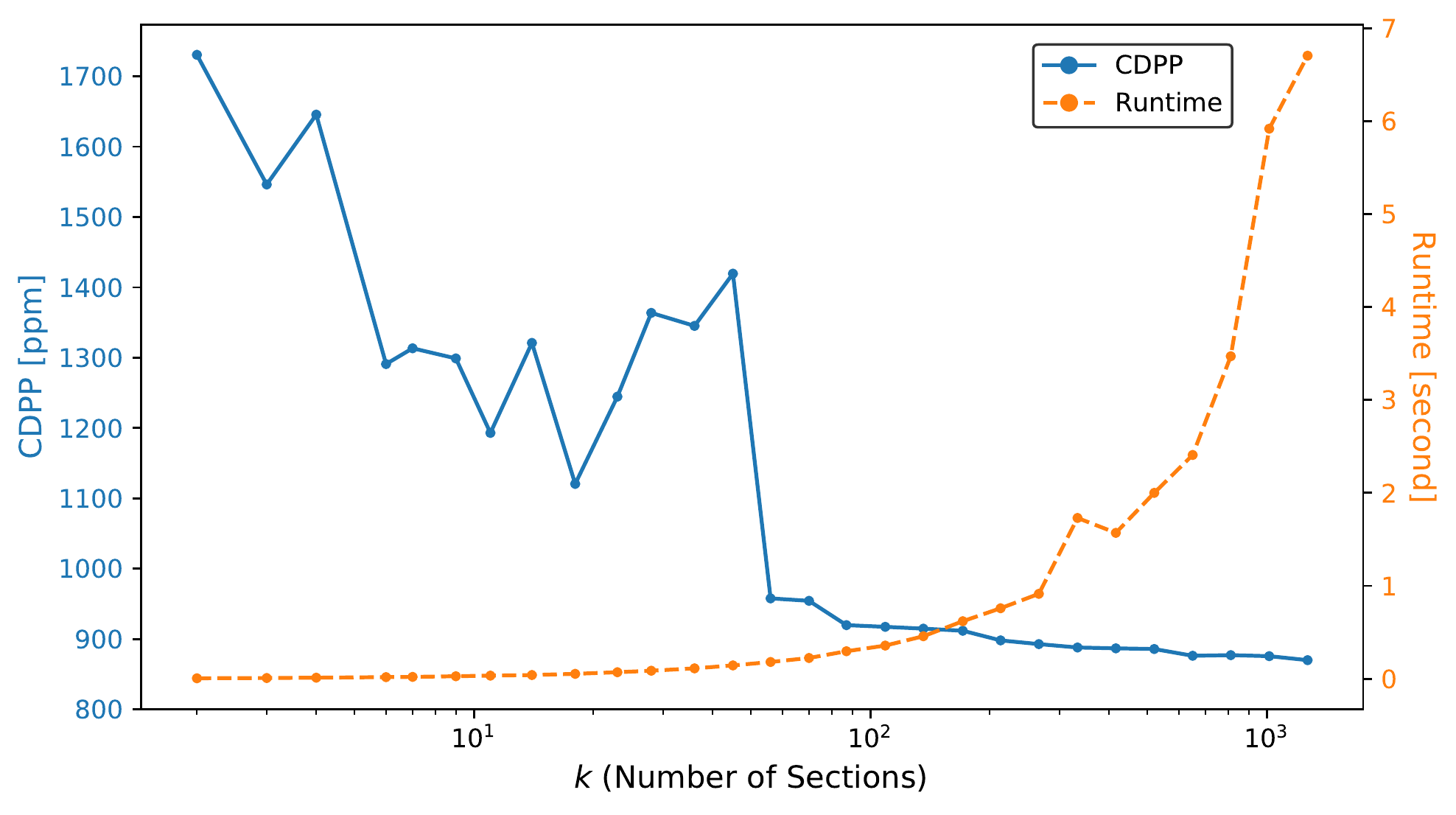}
    \caption{The 6-hr CDPP and runtime for de-trending the central pixel of ASSASSN-18tb as a function of $k$ ($L=256, \lambda_\mathbf{L}=\lambda_\mathbf{P}=10^{-2}$). Note that the x-axis is scaled logarithmically. The 6-hr CDPP fluctuates while decreasing until around $k{\sim}45$ and then shows a significant decrease beyond that point. It then shows marginal decreases from $k{\sim}100$ while the runtime increases roughly linearly.}
    
    \label{fig:k_vs_cdpp_and_executiontime}
\end{figure}

In Figure \ref{fig:comparing_reg_parameters} we show the effect of changing the regularization parameters by running the de-trending method on a coarse grid of $\lambda_\mathbf{L}$ and $\lambda_\mathbf{P}$ and calculating the 6-hr CDPP. Each panel shows the same de-trended pixel light curve, the central pixel for ASASSN-18tb, and the 6-hr CDPP for a different combination of $\lambda_\mathbf{L}$ and $\lambda_\mathbf{P}$ while we kept the other values constant ($L=256, k=126$). The values that both $\lambda_\mathbf{L}$ and $\lambda_\mathbf{P}$ can take are $[0,\; 10^{-3},\; 10^{-2},\; 10^{-1},\; 10^1,\; 10^3]$. The values of $\lambda_\mathbf{L}$ are in increasing order from the most left column ($\lambda_\mathbf{L}=0$) to the most right column ($\lambda_\mathbf{L}=10^3$). The values of $\lambda_\mathbf{P}$ are in increasing order from the top row ($\lambda_\mathbf{P}=0$) to the bottom row ($\lambda_\mathbf{P}=10^3$). In other words, the regularization increases the further right or down the panel is. Looking at the two most right columns, particularly the most right column, we see that more of the systematic effects remain in the de-trended light curve. This behavior is due to the large value of $\lambda_\mathbf{L}$ causing the systematics model to be rigid and therefore underfit. In the lower rows where $\lambda_\mathbf{P}$ is large we see that the polynomial model is underfitting and either the systematics model overfits and removes the long-term trend (in the left columns) or the systematics model also underfits and the de-trended light curve looks similar to the original light curve (in the right columns). The smallest values for the 6-hr CDPP are obtained in the third column ($\lambda_\mathbf{L}=10^{-2}$) where it is ${\sim}904\,\mathrm{ppm}$ in the top four rows. We see that for each of the columns the top four rows show essentially the same results both in terms of the shape of the de-trended light curve and the 6-hr CDPP, indicating that the results are not particularly sensitive to the choice of $\lambda_\mathbf{P}$ and that there is a range of ``good" values (spanning several orders of magnitude) when using a cubic polynomial. We further investigated this behavior by experimenting with higher degree polynomials such as a 13th-degree polynomial and a 50th-degree polynomial, and found that the results become more sensitive to the value of $\lambda_\mathbf{P}$ as these flexible polynomials start to capture the same variations as the systematic model when not sufficiently regularized. 

In Figure \ref{fig:comparing_kvalues} and \ref{fig:k_vs_cdpp_and_executiontime} we show the effect of changing $k$, the number of sections to use for the train-and-test framework \edit1{while keeping the other parameters constant ($L=256, \lambda_\mathbf{L}=\lambda_\mathbf{P}=10^{-2}$)}. As explained in \S\ref{sec:train_and_test_framework}, \textsl{each} section of the light curve is de-trended using the coefficients obtained from fitting to the \edit1{remainder} of the light curve. The smallest possible value is $k=2$ and the largest possible value is when $k$ is equal to the number of data points. Figure \ref{fig:comparing_kvalues} shows the de-trended pixel light curve and the calculated 6-hr CDPP for several different values of $k$. We see that the 6-hr CDPP decreases as we increase $k$ but at the cost of a higher runtime. To further investigate this relationship, we plotted the 6-hr CDPP and the runtime for a given value of $k$ in Figure \ref{fig:k_vs_cdpp_and_executiontime}. The 6-hr CDPP fluctuates while decreasing until around $k{\sim}45$ and then shows a significant decrease beyond that point. It then shows marginal decreases from $k{\sim}100$ while the runtime increases roughly linearly. 
This behavior shows that setting $k$ in the ${\sim}100{-}200$ range is a good trade-off between minimizing the CDPP and the runtime. In general, we believe that this range will be appropriate for most use cases. However, we also recommend the user try a range of values to check for any significant differences between the de-trended light curves. Additionally, for Cycle 3 and beyond where the FFIs will be taken at 10-minute cadence, the value of $k$ will likely need to be increased. 

\section{Additional Examples \& Discussion} \label{sec:results_examples}
In this section we show and discuss additional de-trending results for several sources: the tidal disruption event ASASSN-19bt \citep{holoien2019}, an exoplanet-hosting star TOI-172 (TIC 29857954) \citep{rodriguez2019}, and a fast-rotating variable star (TIC 395130640). When appropriate, we compare our method to other de-trending pipelines.

\subsection{TDE ASASSN-19bt} \label{sec:asassn-19bt}
Tidal disruption events (TDEs) occur when a star passes sufficiently close to a supermassive black hole and is torn apart by the black hole's tidal forces. As some of the star's mass form an accretion disk around the black hole, it produces an observable bright flare. 
ASASSN-19bt/AT 2019ahk,
initially discovered by the ASAS-SN project in January 2019 
, was a TDE observed by TESS in its southern Continuous Viewing Zone (CVZ) \citep{stanek2019, holoien2019}. The CVZ is a ${\sim}12^\circ$ radius circular region of the sky centered at the ecliptic pole, in this case the southern ecliptic pole, that TESS observes for roughly a full year (one Cycle).
As a source located at RA, Dec (J2000):  $105.04811^\circ, -66.04004^\circ$ in the southern CVZ, ASASSN-19bt was observed between Sectors 1-13 excluding Sector 6 where the source fell on a chip gap. The TDE was first observed in the second orbit of Sector 7.

\begin{figure}[t]
    \centering
    \includegraphics[width=\textwidth]{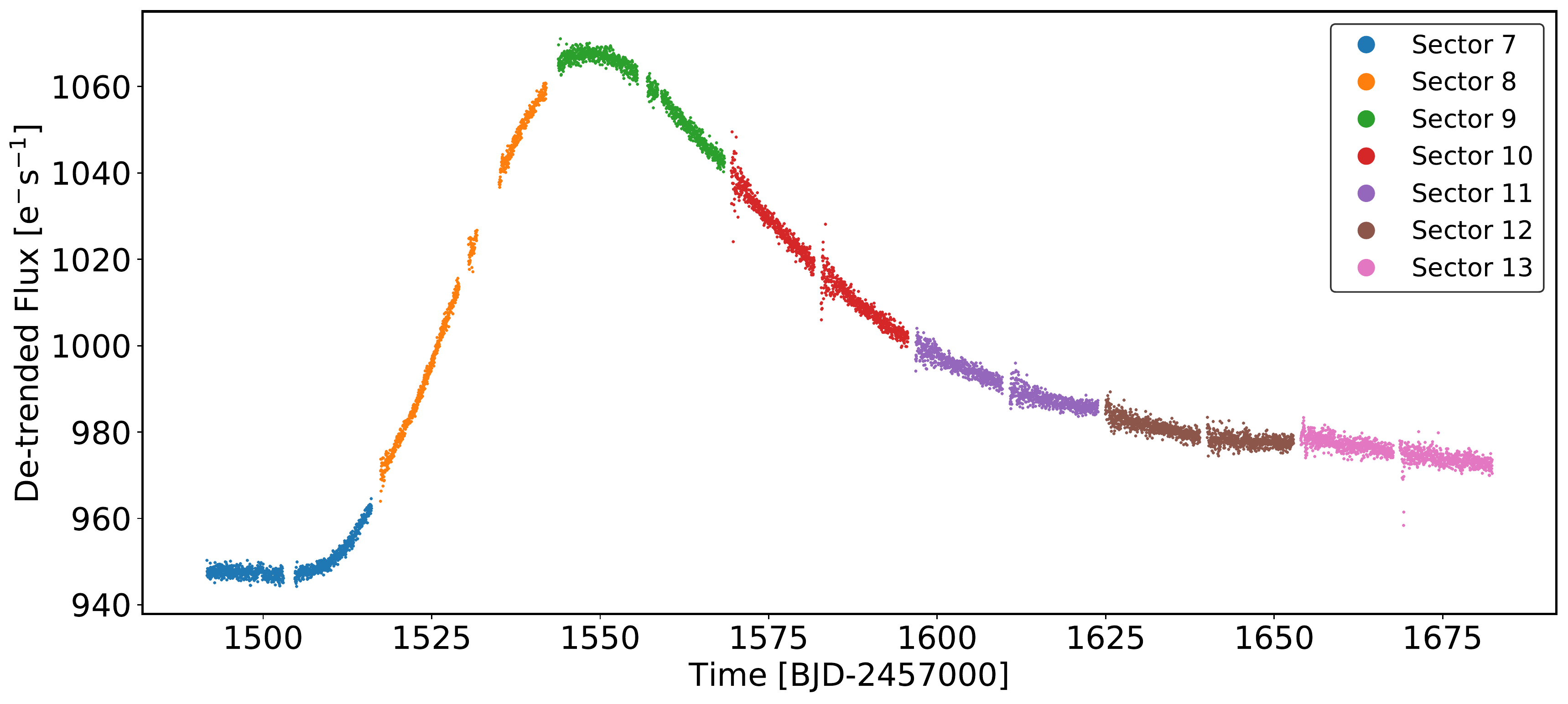}
    \caption{The multi-sector de-trended light curve of the tidal disruption event ASASSN-19bt.
    Each sector was de-trended separately and the de-trended light curves were stitched together as described in the text. For each sector we chose a $3 \times 3$ aperture centered on the source, did not perform an initial background subtraction, added a cubic polynomial, and set $L=256, \lambda_\mathbf{L}=\lambda_\mathbf{P}=0.01, k=150$.}
    \label{fig:asassn-19bt}
\end{figure}

In \ref{fig:asassn-19bt} we show the results of de-trending Sectors 7-13, which is when the TDE occurred. Data from each sector was de-trended separately as the locations of the predictors pixels are not consistent across sectors. We did not perform the initial background subtraction for this source. For each sector we chose a $3 \times 3$ aperture centered on the source, added a cubic polynomial, and used $L=256, \lambda_\mathbf{L}=\lambda_\mathbf{P}=0.01, k=150$ in the de-trending step. \edit1{As our method of rescaling the dimensionless de-trended light curve back to units of flux is not designed to obtain accurate absolute flux measurements, there are offsets between the de-trended light curves. One potential cause of these offsets is that consistent flux measurements of a source across multiple sectors are difficult as the source falls on different pixels each sector, each with a unique unknown response function \citep{montet2017}.}
To correct for these offsets we first take the last 100 points from the Sector 7 light curve and the first 100 points from the Sector 8 light curve\edit1{,} simultaneously fit for the slope and the two average values of these two portions, and then shift the Sector 8 light curve by adding the linear offset and the difference of the two means. This correction therefore shifts the Sector 8 light curve relative to the Sector 7 light curve. After the Sector 7 and Sector 8 are stitched together, we repeat this process with the partially stitched light curve and the remaining sectors' light curves until all sectors are stitched together. While this stitching is currently implemented in the package, it is not ideal as this simple linear model is unable to capture any non-linear offsets. However, as stitching light curves is a separate \edit1{and difficult} problem, it is beyond the scope of this project to optimize this process.
The de-trended light curves from Sectors 7-9 are comparable to those shown in \citet{holoien2019}, where they used the ISIS image subtraction package \citep{alard1998, alard2000} to create difference light curves for these three sectors. To our knowledge, our study is the first time the de-trended light curves have been presented for Sectors 10-13. 

\subsection{TOI-172} \label{sec:toi172}
TOI-172 (TIC 29857954) is a slightly evolved G star hosting a hot Jupiter exoplanet (TOI-172 b) with a 9.48-day orbital period \citep{rodriguez2019}. As the star was not pre-selected as a 2-minute cadence target, TOI-172 b was the first confirmed exoplanet in the TESS FFI data \citep{rodriguez2019}.

\begin{figure}[t!]
    \centering
    \includegraphics[width=\textwidth]{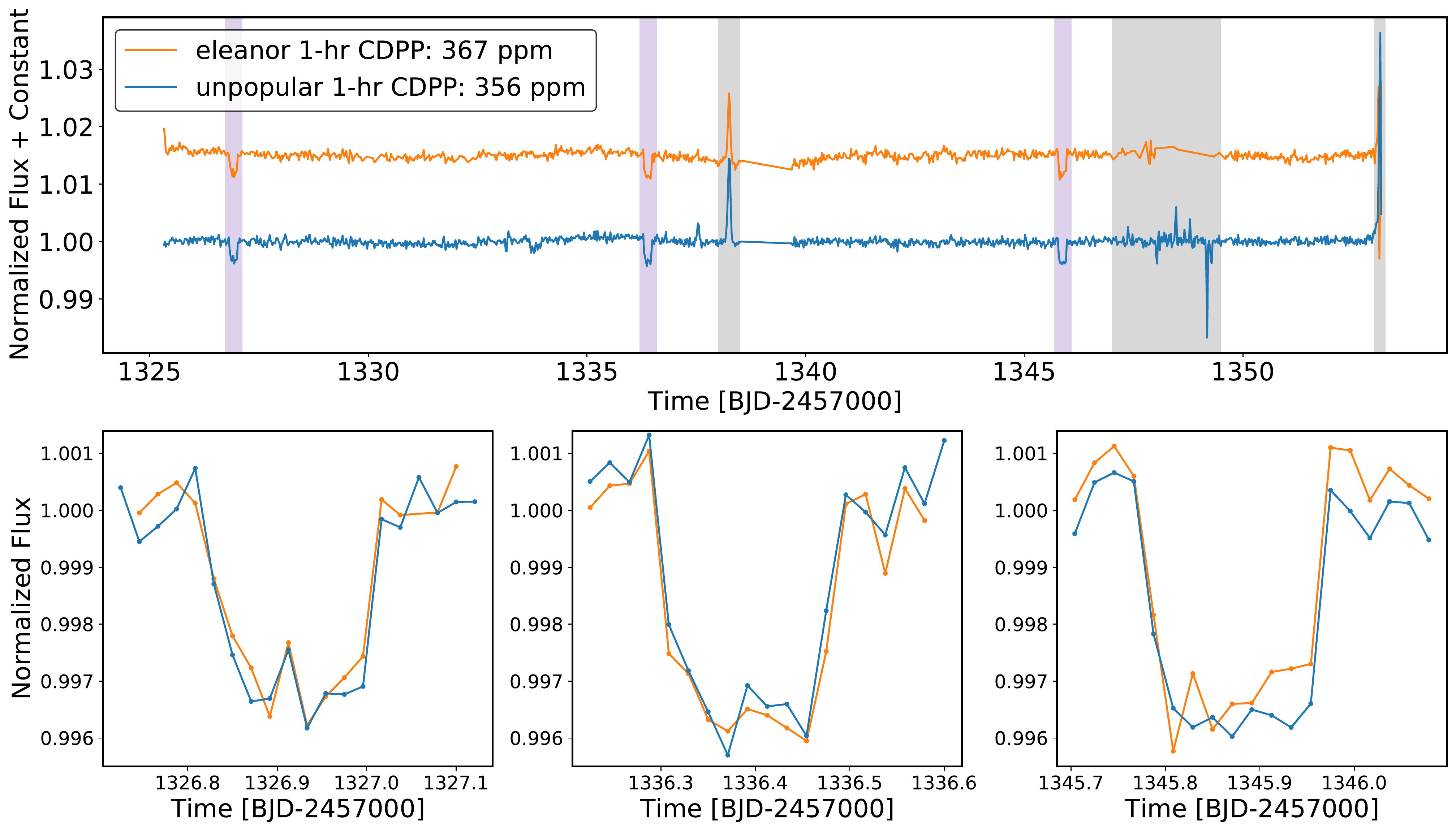}
    \caption{A comparison of de-trending results between the \texttt{eleanor} and \texttt{unpopular} packages. In the top panel we show the two de-trended light curves. Similar to \citet{feinstein2019}, we remove the three transit signals (purple shaded regions) and three specific periods (gray shaded regions), before calculating the 1-hr CDPP using the \texttt{lightkurve} package. The first region is when an asteroid crossed the aperture, the second region is when an improper spacecraft configuration caused high pointing jitter, and the third region is when scattered light from the Moon caused large spikes. The \texttt{unpopular} light curve has a lower 1-hr CDPP of 356 ppm. To obtain the \texttt{unpopular} light curve, we chose a $3 \times 3$ aperture centered on the source, did not add a polynomial component, performed the initial background subtraction, and set $L=256, \lambda_\mathbf{L}=0.01, k=150$. In the bottom three panels we show the de-trended light curves around the three transit signals. The transit depths are comparable between the two light curves.
    } 
    \label{fig:toi172}
\end{figure}

While \texttt{unpopular} was not optimized to obtain light curves with transit signals, we tested our de-trending method on Sector 1 data and present our results in Figure \ref{fig:toi172}. We chose a $3 \times 3$ aperture centered on the source, did not add a polynomial component, performed the initial background subtraction, and used $L=256, \lambda_\mathbf{L}=0.01, k=150$ in our de-trending step. The initial background subtraction was necessary to prevent the transits from becoming shallower in the normalized flux space. After identifying the times where transits occurred, in this case by finding the outliers to the running average of the de-trended light curve, we masked these transits and \textsl{reran the de-trending step on the original data}. This ``masking" approach of removing transits during the de-trending step can prevent our misspecified model\edit1{, as it does not include a transit model,} from erroneously fitting to the transit signals and distorting them. This approach is explained in detail in Section 4.1 of \citet{luger2016}. While the difference was negligible for this source, for light curves containing large signals (e.g., deep transits, eclipsing binaries, flares, microlensing events) this approach will likely result in substantial improvement. \edit1{Therefore, if a user obtains a de-trended light curve showing large signals, we recommend that they mask these points and rerun the de-trending step. Otherwise, it becomes possible for the systematics model to fit to these strong signals in order to minimize the objective function, leading to underfitting to the remainder of the light curve and a distorted de-trended light curve.} An alternative but computationally expensive approach to prevent distorting the transit signal, used in \citet{foreman-mackey2015}, is to fit a joint transit and systematics model. \edit1{On a related note, there may be cases when there are large signals \textsl{in the regressors}. In theory, if a significant fraction of the regressors contain them it becomes possible for these signals to be injected into the de-trended light curve. However, in practice this scenario is unlikely as we choose many regressors and therefore reduce the fraction of signals that contain such signals. A small number of pixels containing such signals would not pose an issue as the regression step would calculate smaller weights for them.}

In the top panel of Figure \ref{fig:toi172} we also show the corrected light curve obtained using the \texttt{eleanor} package \citep{feinstein2019}. Following a similar approach to \citet{feinstein2019}, we calculated the 1-hr CDPP for each light curve after removing the transits and specific excluded periods. The first excluded period $(1338 < \mathrm{Time} < 1338.5)$ is when an asteroid crossed the aperture \citep{rodriguez2019}, the second excluded period $(1347 < \mathrm{Time} < 1349.5)$ is when the spacecraft experienced high pointing jitter due to an improper configuration\footnote{\url{https://archive.stsci.edu/missions/tess/doc/tess_drn/tess_sector_01_drn01_v02.pdf}}, and the third excluded period $(\mathrm{Time} > 1353)$ is when the scattered light from the Moon entered the FOV. The \texttt{eleanor} 1-hr CDPP is 367 ppm and the \texttt{unpopular} 1-hr CDPP is 356 ppm. This minor difference in the CDPP should \textsl{not} be viewed as \texttt{unpopular} outperforming \texttt{eleanor} given how various choices during the de-trending step can easily change the CDPP. In the bottom three panels, we show the de-trended light curves around the three transit signals. The overall shape and the depth between the two light curves is comparable. We note that the initial background subtraction was necessary to recover this transit depth. Omitting the initial background subtraction resulted in shallower transits as the estimated baseline flux was higher, resulting in suppressed depths in the normalized flux space. The comparable performance of  to the \texttt{eleanor} package shows that 
with similar CDPP and recovered transit depths to the \texttt{eleanor} package, this result indicates that \texttt{unpopular} can also be used for studying exoplanets with comparable performance to other packages optimized for exoplanet searches.

\subsection{TIC 395130640} \label{sec:tic395130640}

\begin{figure}[t]
    \centering
    \includegraphics[width=\textwidth]{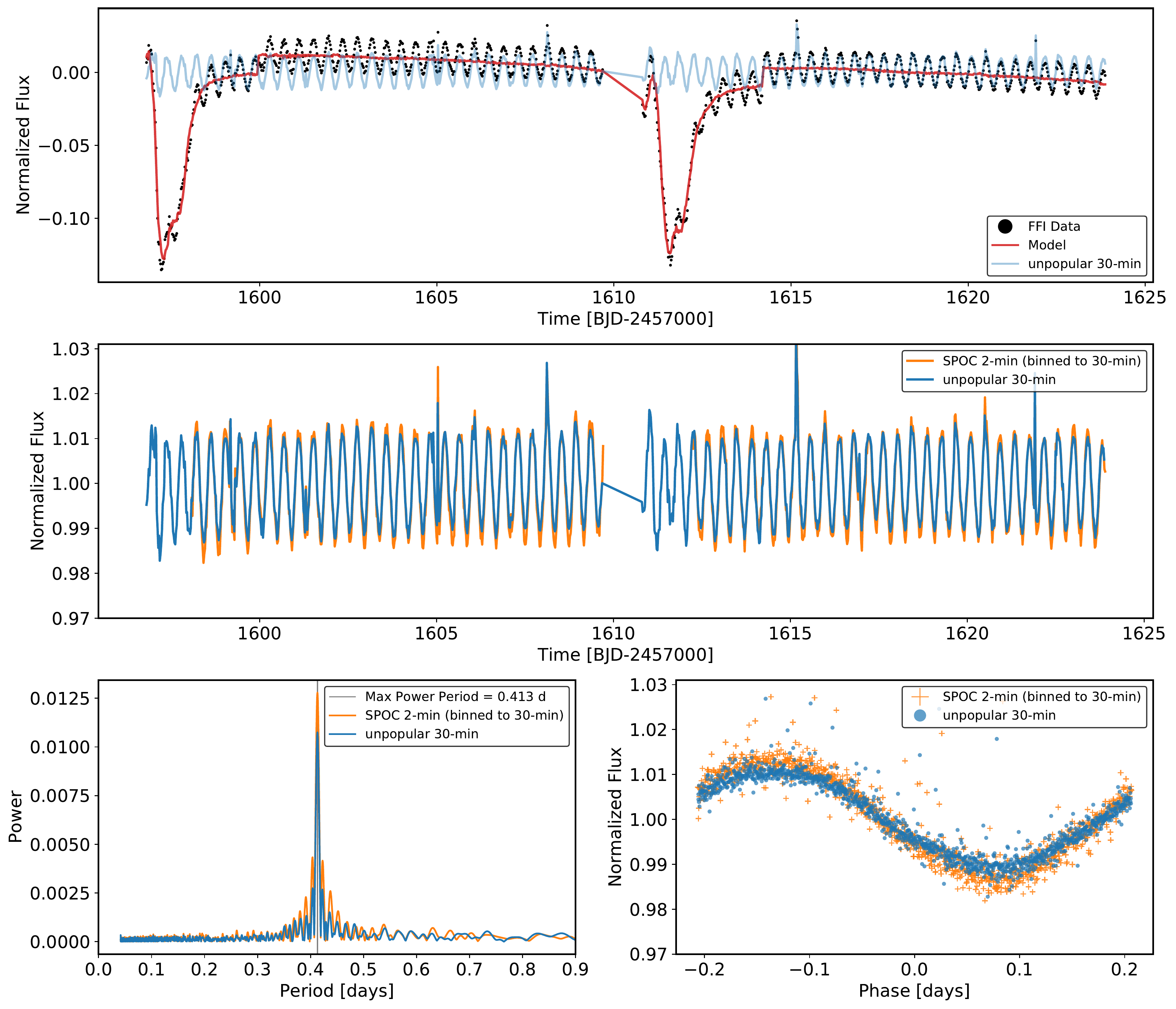}
    \caption{Analysis of Sector 11 TIC 39513064 data. In the top panel we show the background-subtracted and normalized FFI data (black), the systematics prediction from the \texttt{unpopular} package (red), and the de-trended light curve (blue) centered around zero. The two dips at the beginning of each orbit are due to the initial background subtraction overcorrecting the scattered light signal. In the center panel we show the SPOC 2-minute light curve binned to 30 minutes (orange) and the de-trended \texttt{unpopular} FFI light curve (blue). Two regions of strong scattered light at the beginning of each orbit were removed in the SPOC light curve as it affected the SPOC pipeline's systematics removal process. In the bottom-left panel we plot the Lomb-Scargle periodograms for these two light curves, showing that they both peak at 0.413 days. 
    We folded both light curves with this period and show them in the bottom-right panel. To obtain the \texttt{unpopular} light curve, we used the SPOC aperture, performed an initial background subtraction, did not add a polynomial component, and set $L=256, \lambda_\mathbf{L}=1, k=150$.}
    \label{fig:tic39513}
\end{figure}

TIC 395130640 (2MASS J11165730-8027522) is an M dwarf with $M = 0.137\;M_\odot, R = 0.179\;R_\odot$, and a measured rotation period of $P = 0.413\;\mathrm{days}$ \citep{newton2018}. TIC 395130640 was observed in Sectors 11 \& 12, and as a source in the TESS Candidate Target List has 2-minute cadence data de-trended by the SPOC pipeline \citep{jenkins2016} We chose to demonstrate our package's performance on this source simply because it allows us to compare our performance to the official SPOC pipeline and also to verify that we can recover the same rotation period as the previously published value. 

Prior to de-trending the Sector 11 FFI data we performed the initial background subtraction as we saw that that the FFI flux values were ${\sim}30\%$ higher than the \edit1{2-minute cadence Target Pixel File (TPF)} values for large sections of the light curve. After this subtraction the FFI values were only ${\sim}6\%$ higher. Investigating the cause of this difference is beyond the scope of this paper, but it is likely a product of how the SPOC pipeline calibrates the FFI data and TPF data differently. It is possible that this issue can be attributed to the known issue where the SPOC pipeline can overestimate the background flux level and artificially lower the baseline flux level. \citep{kostov2020, feinstein2020, burt2020}

After this initial background subtraction, we de-trended the Sector 11 FFI data with the same aperture as the SPOC pipeline, did not add a polynomial component, and used $L=256, \lambda_\mathbf{L}=1, k=150$ in our de-trending step. \edit1{We set $\lambda_\mathbf{L}=1$ as we saw that, after de-trending with a range of values ($\lambda_\mathbf{L}=[0.01, 0.1, 1, 10, 100]$), the $\lambda_\mathbf{L}=1$ de-trended light curve was the most visually similar to the SPOC light curve. Larger values of $\lambda_\mathbf{L}$ resulted in the de-trended light curves still containing the characteristic scattered light systematics, while smaller values resulted in the de-trended light curve resembling white noise, a potential sign of overfitting}. Note that in this case, as our model does not have a sinusoidal component, there is significant \edit1{model} misspecification, and regularizing the systematics component more strongly was beneficial.

In the top panel of Figure \ref{fig:tic39513} we show the background-subtracted and normalized FFI data, the prediction from the systematics model, and the de-trended light curve obtained by subtracting the systematics model from the normalized FFI data. These light curves are centered about zero as they have not been rescaled to the original flux levels.
The two large dips at the beginning of each orbit in the FFI data are due to the initial background subtraction overcorrecting the scattered light signal. In the center panel we show the median-normalized SPOC 2-minute light curve binned to 30 minutes, to match the cadence of the FFI data, and the median-normalized de-trended \texttt{unpopular} FFI light curve. While the two regions at the beginning of each orbit were removed in the 2-minute data as the strong scattered light from the Earth affected the SPOC pipeline's systematics removal step,\footnote{\url{https://archive.stsci.edu/missions/tess/doc/tess_drn/tess_sector_11_drn16_v02.pdf}} our method was able to de-trend those regions. The \texttt{unpopular} light curve and the binned SPOC light curve show marginal differences besides the smaller amplitude of variability in the \texttt{unpopular} light curve. We believe this amplitude difference is a result of the additional corrections being applied in the PDC module of the SPOC pipeline, as we saw that the amplitudes between the SPOC SAP and the \texttt{unpopular} light curves were similar. While we checked to see whether the difference between the SAP and the PDCSAP flux could be explained by the CROWDSAP value, the ratio of target flux to total flux in the optimal aperture, the differences were still present after multiplying the CROWDSAP value to the PDCSAP flux value.
Regardless of the origin of this discrepancy, as PDCSAP light curves are the standard we opt to compare our light curve to those. We used the \texttt{lightkurve} package to identify the periods of the two light curves using the Lomb-Scargle periodogram \citep{lomb1976, scargle1982, vanderplas2018}. The periodograms are shown in the bottom-left panel of Figure \ref{fig:tic39513}. The maximum power, the semi-amplitude of the oscillation, for the binned SPOC light curve is 0.0128 while it is 0.0107 for the \texttt{unpopular} light curve. The period at maximum power for both light curves is $P=0.413\;\mathrm{days}$ and is in agreement with the previously published value \citep{newton2018}. We folded both light curve with this period of 0.413 days and show them in the bottom-right panel. These results indicate that our lightweight method will allow users to efficiently produce de-trended FFI light curves of comparable quality to the 2-minute light curves produced by the SPOC pipeline. 

\section{Discussion \& Conclusion} \label{sec:conclusion}


We have presented \texttt{unpopular}, an open-source Python package to efficiently extract de-trended light curves from TESS FFIs. The \texttt{unpopular} package is based on the \textsl{CPM} methods, originally developed for Kepler and K2 data \citep{wang2016, wang2017}, and employs regularized linear regression (i.e., ridge regression) to de-trend individual pixel light curves. Using the realization that pixels illuminated by different sources on the same CCD are likely to share similar systematic effects but unlikely to share similar astrophysical trends, the regressors for this data-driven linear model are pixel light curves from other distant (and thus causally disconnected) sources in the image. To prevent overfitting, we incorporated $L_2$ regularization and a train-and-test framework where the data points used to obtain the model coefficients and the data points being de-trended are mutually exclusive. We also allow for adding an additional polynomial component in the model to capture and preserve any long-term trends that may be astrophysical in the data. As the model is linear, de-trending takes only a few seconds for a given source. 

We validated our method by de-trending a variety of sources and comparing them to those obtained by other pipelines. Our de-trended light curves for the supernova ASASSN-18tb and the TDE ASASSN-19bt are similar to the published light curves (ASASSN-18tb: \citet{vallely2019}, ASASSN-19bt: \citet{holoien2019}), where they used the ISIS image subtraction package. We show the de-trended light curves for Sectors 7-13 of ASASSN-19bt, where light curves for Sectors 10-13 have not been published before. For TOI-172, an exoplanet-hosting star, we compared our light curve to that obtained by the \texttt{eleanor} package. Our package was able to recover the transits and also produce a light curve with a lower CDPP than \texttt{eleanor}, indicating that our package can also be used for exoplanet studies. For TIC 395130640, a fast-rotating M dwarf, we compared our FFI light curve to the 2-minute light curve de-trended by the SPOC pipeline. Other than a slightly decreased amplitude in the variation, our light curve shows the same structure as the SPOC light curve, indicating that our lightweight package can produce light curves with similar quality to the SPOC pipeline. Additionally, our package was able to de-trend regions of strong scattered light that the SPOC pipeline was not able to.

While we have presented favorable results, as with any method there are limitations to our approach. One limitation is the lack of an automated method to choose the optimal set of parameters for the de-trending process: the number of predictor pixels $L$, the selection method for choosing the set of predictor pixels, the decision of whether to incorporate a polynomial component, the degree of the polynomial, the regularization parameters $\lambda_\mathbf{L}, \lambda_\mathbf{P}$, and the number of sections to use for the train-and-test framework $k$. While one could in principle make these choices based on minimizing the CDPP, this approach is only appropriate for users interested in exoplanets. For other sources, the user will have to make some of these choices based on the signal they are interested in, such as how the inclusion of the polynomial component and its degree is based on whether the user is attempting to recover sources that contain long-term astrophysical trends (e.g., supernovae, tidal disruption events, slow-rotating stars). Other choices can be made heuristically, such as how the value of $\lambda_\mathbf{L}$, as also mentioned in \S\ref{sec:tic395130640}, will likely need to become larger the more the model is a misspecification for the underlying data. For the examples given in this paper, we chose the values based on visual inspection of the de-trended light curve. Visual inspection is reasonable for sources that already have published light curves, but is impractical for blind searches as the number of light curves will be large. Overcoming this limitation \edit1{for this method and any flexible method with tuning parameters} will likely require the development of various light curve ``goodness" metrics that can be optimized for different types of sources. \edit1{The development of these metrics will be crucial as the community continues to develop flexible data-driven models.}

Another limitation is the use of the polynomial component for capturing trends in the data. While the polynomial is effective at capturing certain long-term trends, it is neither physically motivated nor appropriate for certain variations. For example, the sinusoidal variation of TIC 395130640 cannot be captured with a low-degree polynomial. A sufficiently high-degree polynomial could potentially capture the variation but would also likely start fitting to the systematic effects. One approach, used in \citet{angus2016} and \citet{hedges2020} is to explicitly include a sinusoidal component and simultaneously fit the systematic effects and periodic variations over a grid of frequencies. Another slightly more general approach, employed in the \texttt{EVEREST} and \texttt{EVEREST 2.0} pipelines, is to replace the polynomial component with a Gaussian process. While both approaches are more computationally expensive, there will likely be improvements depending on the source.

While the method takes only a few seconds to de-trend a given source, we could still reduce the computational time by parallelizing the code. As each pixel is de-trended independently, parallelizing the code is conceptually straightforward. Significantly reducing the computational time of this method could allow for the production of high-level science products where we de-trend every pixel in the TESS FFIs. These de-trended FFIs could be helpful for detecting both transient events and moving sources in an FFI.

The code is open source under the MIT license and available at \url{https://github.com/soichiro-hattori/unpopular} along with a Jupyter notebook tutorial. We hope that this method and the techniques used here can be generalized and applied to upcoming missions such as \textsl{PLATO} \citep{rauer2014} and the LSST survey at the Vera C. Rubin Observatory \citep{ivezic2019} in removing systematic effects to make new discoveries.

\acknowledgments
SH would like to thank Joseph D. Gelfand (NYUAD) for his support during this project. We also thank the reviewer for thorough and constructive feedback that greatly improved this paper.

This project was developed in part at the Expanding the Science of TESS meeting, which took place in 2020 February at the University of Sydney and also at the online.tess.science meeting, which took place globally in 2020 September. SH was supported by the NYU Abu Dhabi Research Enhancement Fund (REF) under grant RE022. RA acknowledges support by NASA under award numbers 80NSSC21K0096 and 80NSSC21K0636. TP acknowledges support from NASA under the Swift GI grant 1619152, the Tess GI grant G03267, and support from the the NYU Center for Cosmology and Particle Physics. This paper includes data collected by the TESS mission. Funding for the TESS mission is provided by the NASA's Science Mission Directorate. This research has made use of NASA's Astrophysics Data System Bibliographic Services. 
%

\vspace{5mm}
\facilities{TESS}


\software{Astrocut \citep{brasseur2019}, Astropy \citep{astropy2013, astropy2018}, eleanor \citep{feinstein2019}, lightkurve \citep{barentsen2020}, Matplotlib \citep{hunter2007}, NumPy \citep{harris2020}, Scikit-learn \citep{pedregosa2011}, SciPy \citep{virtanen2020}}



\appendix
\section{Solutions to Linear Least Squares} \label{appendix:proof_ls}
Consistent with the notation in \S\ref{sec:cpm}, lowercase upright fonts in bold ($\mathbf{y}$) are row or column vectors and uppercase upright fonts in bold ($\mathbf{X}$) are matrices. 

\subsection{Ordinary Least Squares} \label{appendix:proof_ols}
Here we derive the set of coefficients $\mathbf{\hat{w}}$ that minimizes the sum of squared residuals

\begin{equation}
    S(\mathbf{w}) = \mathbf{\lVert y - X \cdot w \rVert_\mathrm{2}^\mathrm{2} = (y - X \cdot w)^\top \cdot (y - X \cdot w)}.
\end{equation}
Expanding this expression gives
\begin{align}
    S(\mathbf{w}) & = \mathbf{(y - X \cdot w)^\top \cdot (y - X \cdot w)} \\ 
    & = \mathbf{(y^\top - w^\top \cdot X^\top) \cdot (y - X \cdot w)} \\ 
    & = \mathbf{(y^\top \cdot y) - (y^\top \cdot X \cdot w) - (w^\top \cdot X^\top \cdot y) + (w^\top \cdot X^\top \cdot X \cdot w)} \\ 
    S(\mathbf{w}) & = \mathbf{(y^\top \cdot y) - \mathrm{2}(w^\top \cdot X^\top \cdot y) + (w^\top \cdot X^\top \cdot X \cdot w)}
\end{align}
where we use $\mathbf{y^\top \cdot X \cdot w = (y^\top \cdot X \cdot w)^\top = w^\top \cdot X^\top \cdot y}$ as it is a scalar. To find the $\mathbf{w}$ that minimizes $S(\mathbf{w})$, we take the derivative of $S(\mathbf{w})$ with respect to $\mathbf{w}$ and set it to zero. Using the denominator layout convention for matrix differentiation, 
\begin{align}
    \frac{\partial S(\mathbf{w})}{\partial \mathbf{w}} & = \mathbf{\frac{\partial}{\partial w}(y^\top \cdot y) - \mathrm{2}\frac{\partial}{\partial w}(w^\top \cdot X^\top \cdot y) + \frac{\partial}{\partial w}(w^\top \cdot X^\top \cdot X \cdot w)} \\ 
    & = \mathbf{\mathrm{0} - \mathrm{2}(X^\top \cdot y) + \mathrm{2}(X^\top \cdot X \cdot w)} \\ 
    \frac{\partial S(\mathbf{w})}{\partial \mathbf{w}} & = \mathbf{\mathrm{2}(X^\top \cdot X \cdot w) - \mathrm{2}(X^\top \cdot y) = \mathrm{0} \Rightarrow X^\top \cdot X \cdot w = X^\top \cdot y}.
\end{align}
If $(\mathbf{X^\top \cdot X})$ is an invertible square matrix, we can left multiply $(\mathbf{X^\top \cdot X})^{-1}$ to both sides of the above expression to solve for $\mathbf{\hat{w}}$
\begin{equation}
    \mathbf{\hat{w} = (X^\top \cdot X)^{-1} \cdot (X^\top \cdot y)}.
\end{equation}


\subsection{Ridge Regression} \label{appendix:ridge_regression}
Here we derive the ridge coefficients $\mathbf{\hat{w}_\mathrm{ridge}}$ that minimize the objective function
\begin{equation}
    S_{\mathrm{ridge}}(\mathbf{w}) = \mathbf{\lVert y_*^{} - X_*^{} \cdot w \rVert_\mathrm{2}^\mathrm{2} + \lVert \Gamma \cdot w \rVert_\mathrm{2}^\mathrm{2}}.
\end{equation}

The derivation is similar to that of OLS.
\begin{align}
    S_{\mathrm{ridge}}(\mathbf{w}) ={}& \mathbf{(y_*^{} - X_*^{} \cdot w)^\top \cdot (y_*^{} - X_*^{} \cdot w) + (\Gamma \cdot w)^\top \cdot (\Gamma \cdot w)} \\
    ={}& \mathbf{(y_*^\top - w^\top \cdot X_*^\top) \cdot (y_*^{} - X_*^{} \cdot w)} + \mathbf{(w^\top \cdot \Gamma^\top) \cdot (\Gamma \cdot w)} \\
    \begin{split}
    S_{\mathrm{ridge}}(\mathbf{w}) ={}& \mathbf{(y_*^\top \cdot y_*^\top)} - \mathbf{\mathrm{2}(w^\top \cdot X_*^\top \cdot y_*^{})}\\
    & + \mathbf{(w^\top \cdot X_*^\top \cdot X_*^\top \cdot w)} + \mathbf{(w^\top \cdot \Gamma^\top \cdot \Gamma \cdot w)}
    \end{split}
\end{align}

Taking the derivative of $S_{\mathrm{ridge}}$ and setting it to zero
\begin{align}
    \begin{split}
        \frac{\partial S_\mathrm{ridge}(\mathbf{w})}{\partial \mathbf{w}} ={}& \mathbf{\frac{\partial}{\partial w}(y_*^\top \cdot y_*^\top) - \mathrm{2}\frac{\partial}{\partial w}(w^\top \cdot X_*^\top \cdot y_*^{})}\\
        & + \mathbf{\frac{\partial}{\partial w}(w^\top \cdot X_*^\top \cdot X_*^{} \cdot w) + \frac{\partial}{\partial w}(w^\top \cdot \Gamma^\top \cdot \Gamma \cdot w)}
    \end{split}\\
    ={}& \mathbf{-\mathrm{2}(X_*^\top \cdot y_*^{}) + \mathrm{2}(X_*^\top \cdot X_*^\top \cdot w) + \mathrm{2}(\Gamma^\top \cdot \Gamma \cdot w)}\\
    \begin{split}
    \frac{\partial S_\mathrm{ridge}(\mathbf{w})}{\partial \mathbf{w}} ={}& \mathbf{-\mathrm{2}(X_*^\top \cdot y_*^{})} +  \mathbf{\mathrm{2}(X_*^\top \cdot X_*^\top + \Gamma^\top \cdot \Gamma)\cdot w = \mathrm{0}}\\ 
    \Rightarrow & \mathbf{(X_*^\top \cdot X_*^{} + \Gamma^\top \cdot \Gamma) \cdot w = X_*^\top \cdot y_*}.
    \end{split}
\end{align}
As $\mathbf{(X_*^\top \cdot X_*^{} + \Gamma^\top \cdot \Gamma)}$ is an invertible square matrix, we left multiply its inverse to the above expression to calculate $\mathbf{\hat{w}_\mathrm{ridge}}$
\begin{equation}
    \mathbf{\hat{w}_\mathrm{ridge}} = (\mathbf{X_*^\top \cdot X_*^{}} + \mathbf{\Gamma^\top \cdot \Gamma})^{-1} \cdot (\mathbf{X_*^\top \cdot y_*^{}}).
\end{equation}

\bibliography{ms}{}
\bibliographystyle{aasjournal}


\listofchanges

\end{document}